\documentclass[final,3p,times,authoryear]{elsarticle}

\usepackage{bm}
\usepackage{amsthm, amssymb}
\usepackage{mathtools, cuted}
\usepackage{color}
\usepackage{multicol}
\usepackage[hidelinks]{hyperref}

\newcommand\scalemath[2]{\scalebox{#1}{\mbox{\ensuremath{\displaystyle #2}}}}
\newcommand{\abs}[1]{\vert {#1} \vert}
\newcommand{\dbtilde}[1]{\tilde{\raisebox{2pt}[0.85\height]{$\tilde{#1}$}}}
\newcommand{\average}[1]{\mathrm{E}\left[{#1} \right]}
\newcommand{\std}[1]{\mathrm{std}\left[{#1} \right]}
\newcommand{\Var}[1]{\mathrm{Var}\left[{#1} \right]}
\newcommand{\modif}[1]{\textcolor{black}{#1}}
\newcommand{\review}[1]{\textcolor{black}{#1}}
\newcommand{\KI}{K_\mathrm{I}}
\newcommand{\KIc}{K_\mathrm{Ic}}

\usepackage{lineno}
\journal{International Journal of Solids and Structures}

\begin{document}
	\title{Size effects in the toughening of brittle materials by heterogeneities: a non-linear analysis of front deformations}
	\author[add1]{Mathias Lebihain}
	\author[add2]{Manish Vasoya}
	\author[add3]{V{\'e}ronique Lazarus}
	\cortext[cor1]{Corresponding author : mathias.lebihain@enpc.fr}
	\address[add1]{Navier, Ecole des Ponts, Univ Gustave Eiffel, CNRS, Marne-la-Vall{\'e}e, France}
	\address[add2]{Department of Aerospace Engineering, Texas A\&M University, College Station, TX 77843, USA}
	\address[add3]{IMSIA, ENSTA Paris, CNRS, EDF, CEA, Institut Polytechnique de Paris, Palaiseau, France}
	
	\begin{abstract}
		Traditional computational approaches in simulating crack propagation in perfectly brittle materials rely on the estimate of stress intensity factors along the rupture front. This proves highly challenging in 3D when the crack geometry departs from very specific cases for which analytical solutions are available, like e.g. the penny-shaped crack geometry.
		Here, we extend the first-order theory of \cite{gao_somewhat_1987}, and predict the distribution of the mode I stress intensity factor $\KI$ along the front of a tensile coplanar crack that is slightly perturbed from a reference penny-shaped configuration, up to second order in the perturbation amplitude. 
		Our theory is validated against analytical solutions available for embedded elliptical cracks, and its range of validity is further assessed using numerical simulations performed on cosine front perturbations of varying mode and amplitude. 
		It is then used to develop a homogenization framework for the toughness of weakly disordered media. The effective toughness and its fluctuations are bridged quantitatively to the intensity of the toughness fluctuations and their spatial structure. Our theoretical predictions are compared to the results of $\sim 1$ million simulations of crack propagation building on our second-order theory and Fast Fourier Transforms. We show that the impact of toughness heterogeneities is size-dependent, as they generally weaken the material when the crack size is lower or comparable to the typical heterogeneity size, but reinforces it otherwise. It results in an apparent R-curve behavior of the brittle composite at the macroscale.
	\end{abstract}
	
	\begin{highlights}
		\item Second-order variation of the SIF for small perturbations of a circular crack is derived.
		\item It is used to get the effective toughness of heterogeneous materials \review{in the weak pinning regime}.
		\item This differs from the average of the toughness distribution, predicted by the linear theory.
		\item The effective toughness depends on the crack versus heterogeneity size.
		\item From weaker than the average for small cracks, it becomes stronger for large ones, suggesting a R-curve behavior.
		
	\end{highlights}
	
	\maketitle
	
	\section{Introduction}
	\label{sec:introduction}
	
	Despite the rise of powerful numerical methods, like e.g. cohesive zone models (CZMs) \citep{barenblatt_processzone_1962, xu_numerical_1994, camacho_computational_1996} or phase-field models (PFMs) \citep{francfort_revisiting_1998, hakim_crack_2005, bourdin_variational_2008} that can compute spontaneous three-dimensional crack initiation and propagation in complex structures and multiphysics environments, more ``traditional'' simulation techniques based on the Linear Elastic Fracture Mechanics (LEFM) like e.g. the eXtended Finite Element Method (XFEM) \citep{moes_finite_1999}, have still some value in modeling crack propagation in elaborate settings \citep{paul_3d_2018, lebihain_effective_2021}. This is particularly true when dealing with sharp discontinuities of material properties, as the characteristic length scale associated with the failure process in CZMs and PFMs may interact with that of the fluctuations. In that case, numerical simulations may fail in reproducing the predictions of LEFM, even in simple situations of crack deflection at a bi-material interface \citep{henry_limitations_2019}, and \textit{ah-hoc} numerical compensation methods are required \citep{hansen-dorr_phase-field_2020}. Meanwhile, LEFM-based methods, which build on the \emph{small-scale yielding} assumption, do not suffer these limitations due to the innate localized nature of the dissipation \citep{lebihain_effective_2020}. Yet, in these models, crack extension is not spontaneous, but must be predicted from the combination of \textit{ad-hoc} propagation \citep{griffith_phenomena_1921, irwin_fracture_1958} and direction \citep{erdogan_crack_1963, hussain_strain_1974, goldstein_brittle_1974} criteria. They often require the \textit{a priori} knowledge of the stress intensity factors (SIFs) along the whole crack front. Stress intensity factors are known analytically for a limited number of simple crack geometries and loading conditions \citep{tada_stress_2000}, and they can be computed numerically from e.g. the J-integral \citep{rice_path_1968}, or the virtual crack extension (also called G-theta) method \citep{destuynder_Gtheta_1981, delorenzi_energy_1982}. However, when a crack propagates in a 3D composite material, its front is locally pinned by material heterogeneities, and adopts a tortuous shape that departs from ```standard'' geometries \citep{gao_trapping_1989}. As a result, the SIF distribution along the front cannot be inferred from exact analytical solutions, and the computational cost of SIF evaluation often proves too high to model 3D propagation from the scale of the smallest heterogeneity to that of the structure.
	
	A good compromise can be found by resorting to perturbative approaches of LEFM \citep{lazarus_review_2011}, in which the stress intensity factors along a front that is slightly distorted within its plane from a reference configuration can be estimated from the knowledge of that in the unperturbed configuration. Building on Bueckner-Rice's weight function theory \citep{bueckner_weight_1987, rice_weight_1989}, \cite{rice_first-order_1985} derived the mode I stress intensity factor distribution along a semi-infinite planar crack whose front is slightly perturbed from its reference straight configuration, up to first order in the perturbation amplitude. The approach was then extended to (i) mixed mode conditions by \cite{gao_shear_1986}, (ii) other crack geometries, like e.g. the penny-shaped crack \citep{gao_somewhat_1987, borodachev_variational_2007}, the circular connection \citep{gao_nearly_1987}, tunnel crack \citep{leblond_tensile_1996, lazarus_stability_2002, lazarus_kernel_2002}, interaction of two tunnel cracks \citep{pindra_slit_2010, legrand_slit_2010}, the crack lying between two plates \citep{legrand_plate_2011}, (iii) out-of-plane perturbations \citep{movchan_perturbations_1998, leblond_theoretical_2011}, (iv) dynamic ruptures \citep{perrin_disordering_1994, ramanathan_dynamics_1997}, and (v) cohesive materials \citep{lebihain_cohesive_2022}. The unparalleled computational performances offered by the perturbative methods, through the use of the Fast Fourier Transform (FFT), allowed to investigate the propagation of a crack front in heterogeneous media \citep{gao_trapping_1989}, its roughening under the action of material disorder \citep{schmittbuhl_pinning_1995, rosso_roughness_2002}, and the intermittent statistics that emerge from its interaction with material heterogeneities \citep{ponson_crack_2010, laurson_avalanches_2010}. It has given quantitative means to rationalize the crack-inclusion interactions observed during failure experiments in patterned \citep{bower_bridging_1991, dalmas_crack_2009, chopin_crack_2011, patinet_pinning_2013} and disordered materials \citep{delaplace_high_1999, bares_aftershock_2018}. It has also provided basic ingredients to formulate a homogenization framework for coplanar \citep{roux_effective_2003, patinet_quantitative_2013, demery_geometry_2014, demery_microstructural_2014} and non-coplanar \citep{lebihain_effective_2021} brittle failure. Note that perturbative approaches have also been used in modeling contact and adhesion along homogeneous and heterogeneous surfaces \citep{adda-bedia_instability_2006, xia_toughening_2012, budzik_perturbation_2014, xia_adhesion_2015, argatov_controlling_2021, sanner_crack-front_2022}.
	
	However, the experimental study of \cite{vasoya_experimental_2016} suggests that \emph{higher-order} theories may help to reconcile theoretical predictions with experimental observations of crack front deformation. Moreover, material toughening by heterogeneities has been shown to be of \emph{second-order} in the amplitude of the toughness fluctuations \citep{patinet_quantitative_2013, demery_microstructural_2014}, which is of the same order as the terms omitted by these authors in their asymptotic expansion of the stress intensity factor. As a result, their models only draw conclusions on how material toughening scales with the amplitude of the fluctuations, but they may fail in predicting it quantitatively. Higher-order theories for the crack front perturbations are thus crucial to improve our understanding of the interactions between cracks and heterogeneities. Yet, up to this day, second-order perturbative approaches have only been derived for the semi-infinite crack \citep{adda-bedia_second-order_2006, leblond_second_2012, vasoya_second_2013, vasoya_experimental_2016} loaded in tensile mode I. Their extension to other geometries (like e.g. the penny-shaped crack or the circular connection) may bring valuable insights on a wide variety of physical problems, like e.g. the onset of the fingering instability occurring during crack propagation \citep{vasoya_fingering_2016} or during the contact between an indenter and soft elastic films \citep{yu_adhesion-induced_2021}, the adhesion hysteresis observed along chemically heterogeneous interfaces \citep{sanner_crack-front_2022}, or even the shape of fluid-driven shear ruptures along frictional interfaces governed by Coulomb's friction \citep{saez_three-dimensional_2022}.\\
	
	In this work, we extend the linear model of \cite{gao_somewhat_1987} for the penny-shaped crack loaded in tensile mode I to second-order, following the reasoning of \cite{leblond_second_2012}. We build next on our non-linear theory to investigate material reinforcement arising from the presence of randomly arranged heterogeneities of toughness, and focus in particular on the size effects emerging from the finiteness of the penny-shaped crack geometry.
	
	The paper is organized as follows: in Section 2, we derive a non-linear theory that predicts the mode I stress intensity factor variations arising from coplanar perturbations of the crack front from its reference circular configuration, up to second-order in the perturbation amplitude. This is performed by deriving first the linear expansion of the fundamental kernel associated to the penny-shaped crack geometry. In Section 3, we validate our theory by comparing its predictions with analytical solutions obtained for elliptical cracks loaded by a uniform tensile stress. The validity range of the second-order model is then assessed, building on numerical SIF estimates along wavy crack fronts obtained by the method of \cite{lazarus_numerics_2003} and \cite{david_key_2022} that computes the SIF variations at all orders. Section 4 is devoted to the analysis of quasi-static crack propagation in random fields of toughness, with a particular focus on the toughening arising from these fluctuations. Namely, we discuss how (i) accounting for non-linear terms in the SIF expansion and (ii) the finiteness of the penny-shaped crack geometry may change our results with respect to those available in the literature for the semi-infinite crack \citep{patinet_quantitative_2013, demery_geometry_2014}.
	
	\section{Derivation of the second-order theory}
	\label{sec:theory}
	
	\subsection{Generalities}
	
	We consider a planar penny-shaped crack embedded in an infinite medium made of some isotropically linearly elastic material of Young's modulus $E$ and Poisson's ratio $\nu$. The crack front $\mathcal{F}_0$ describes a circle of radius $a_0$ centered on a point $O$. The crack is contained within the $xOz$ plane. The pair $(r, \theta)$ denotes the coordinates of a point of observation $M$ in the polar coordinate system centered in $O$, and $(\mathbf{e_r}, \mathbf{e_\theta})$ the polar coordinate basis vectors at this point. The crack is loaded in pure mode I through some system of forces, giving rise to a distribution $\KI^0(a_0, \theta)$ of the unperturbed stress intensity factor along the reference circular crack front $\mathcal{F}_0$.
	
\begin{figure}
		\centering
		\noindent\includegraphics[width=0.35\textwidth]{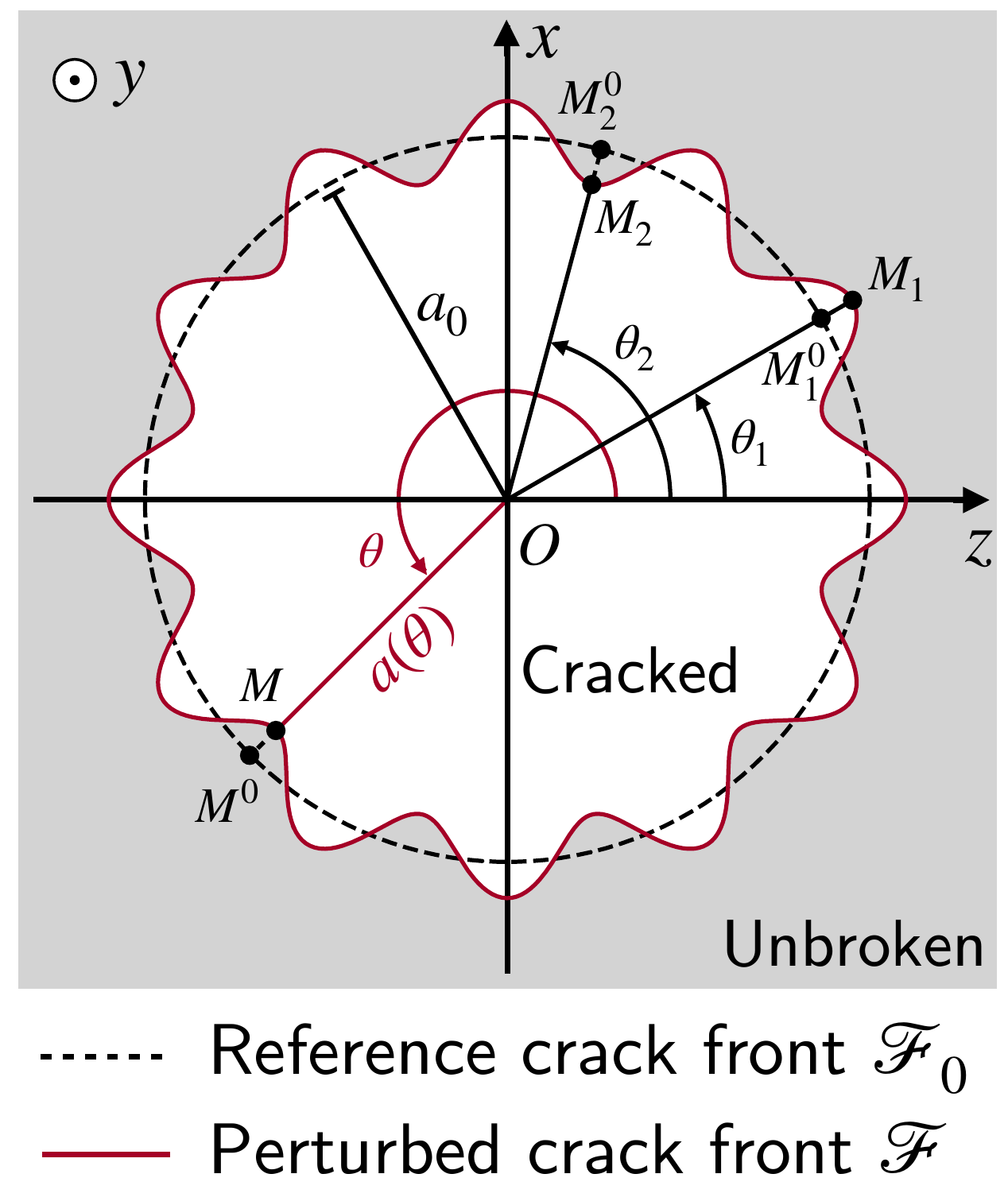}
		\caption{The reference penny-shaped crack with a circular front $\mathcal{F}_0$ of radius $a_0$ (in dashed black line) is perturbed within its plane $(xOz)$ by a small amount $\delta a(\theta)$ in the direction $\mathbf{e_r}$, giving rise to a perturbed crack front $\mathcal{F}$ parametrized by $a(\theta) = a_0 + \delta a(\theta)$ (in solid red line). The points $M_1$, $M_2$, and $M$, indexed by $\theta_1$, $\theta_2$ and $\theta$ respectively, are located on the perturbed front $\mathcal{F}$. $M^0_1$, $M^0_2$ and $M^0$ are their orthogonal projections on $\mathcal{F}_0$.}
		\label{fig:perturbed_crack}
	\end{figure}

	The crack front is now perturbed within its plane in the direction $\mathbf{e_r}$ normal to its reference configuration by a small amount:
	\begin{equation}
		\label{eq:perturbation_shape}
		\delta a(\theta) = A\phi(\theta),
	\end{equation} 
	where $A$ is an infinitesimally small parameter quantifying the non-circularity of the perturbed front $\mathcal{F}$, and $\phi$ is a shape function (see Fig.~\ref{fig:perturbed_crack}). $\delta \KI$ denotes the variations of the mode I SIF arising from the perturbation $\delta a$. Evaluated at a point $M_1$ of $\mathcal{F}$ indexed by $\theta_1$, $\delta \KI$ reads \citep{panasyuk_propagation_1962, gao_somewhat_1987}:
	\begin{equation}
		\label{eq:perturbed_SIF_1stOrder}
		\delta \KI (\theta_1) = \dfrac{\partial \KI^0}{\partial a}(a_0, \theta_1) \delta a(\theta_1) + \mathrm{PV}\int_{0}^{2\pi} Z^0(a_0; \theta_1, \theta) \KI^0(a_0, \theta) \left[\delta a(\theta) - \delta a(\theta_1)\right] a_0 d\theta + \mathcal{O}(A^2),
	\end{equation}
	at first order in the perturbation $\delta a$. Here, the symbol $\mathrm{PV}$ denotes a Cauchy principal value, and $Z^0(a_0; \theta_1, \theta_2)$ is the \emph{kernel} associated with the reference circular crack of radius $a_0$, evaluated at any points $M^0_1$ and $M^0_2$ of $\mathcal{F}^0$, indexed by $\theta_1$ and $\theta_2$ respectively (see Fig.~\ref{fig:perturbed_crack}). It takes the general form \citep{rice_weight_1989}:
	\begin{equation}
		\label{eq:fundamental_kernel}
		Z^0(a_0; \theta_1, \theta_2) = \dfrac{1}{2\pi} \dfrac{W^0(a_0; \theta_1, \theta_2)}{\mathcal{D}(M^0_1, M^0_2)^2},
	\end{equation}
	where $\mathcal{D}(M^0_1, M^0_2)$ is the Euclidean distance between the points $M^0_1$ and $M^0_2$, and $W^0$ is the \emph{fundamental kernel}\footnote{Note the subtle difference of vocabulary: \emph{fundamental kernel} standing for the dimensionless quantity $W$ and \emph{kernel} for $Z$ appearing as the kernel of the integral in Eq. \ref{eq:perturbed_SIF_1stOrder}.} associated with the reference crack front geometry. For a circular crack front, $W^0(a_0; \theta_1, \theta_2)=1$, so that $Z^0$ reads:
	\begin{equation}
		\label{eq:kernel_circular}
		Z^0(a_0; \theta_1, \theta_2) =\dfrac{1}{8\pi a_0^2\sin^2\left[(\theta_2-\theta_1)/2\right]}.
	\end{equation}
	Our goal here is to go beyond the linear expansion~\eqref{eq:perturbed_SIF_1stOrder} of \cite{gao_somewhat_1987}, and derive the expression of the perturbed SIF $\KI$ at \emph{second order} in the perturbation $\delta a$. We follow here the approach of \cite{leblond_second_2012}, and calculate first the variations of the kernels $W$ and $Z$ at \emph{first order} in $\delta a$.
	
	\subsection{First-order expansion of the fundamental kernel}
	
	For tensile cracks, the first-order expansion of the fundamental kernel $W$ has been derived by \cite{rice_weight_1989}. Provided that the perturbation $\delta a$ vanishes at $M_1$ and $M_2$ indexed by $\theta_1$ and $\theta_2$, the variations $\delta W$ of the fundamental kernel write as:
	\begin{equation}
		\label{eq:perturbated_fundamental_kernel_raw}
		\dfrac{\delta W(\theta_1, \theta_2)}{\mathcal{D}(M_1, M_2)^2} = \dfrac{1}{2\pi}\mathrm{PV}\int_{0}^{2\pi} \dfrac{W^0(a_0; \theta_1, \theta)}{\mathcal{D}(M_1, M^0)^2} \dfrac{W^0(a_0; \theta_2, \theta)}{\mathcal{D}(M_2, M^0)^2} \delta a(\theta)\, a_0 d\theta + \mathcal{O}(A^2),
	\end{equation}
	where $M^0$ is the point indexed by $\theta$ on $\mathcal{F}_0$ .
	
\begin{figure}
		\centering
		\noindent\includegraphics[width=0.35\textwidth]{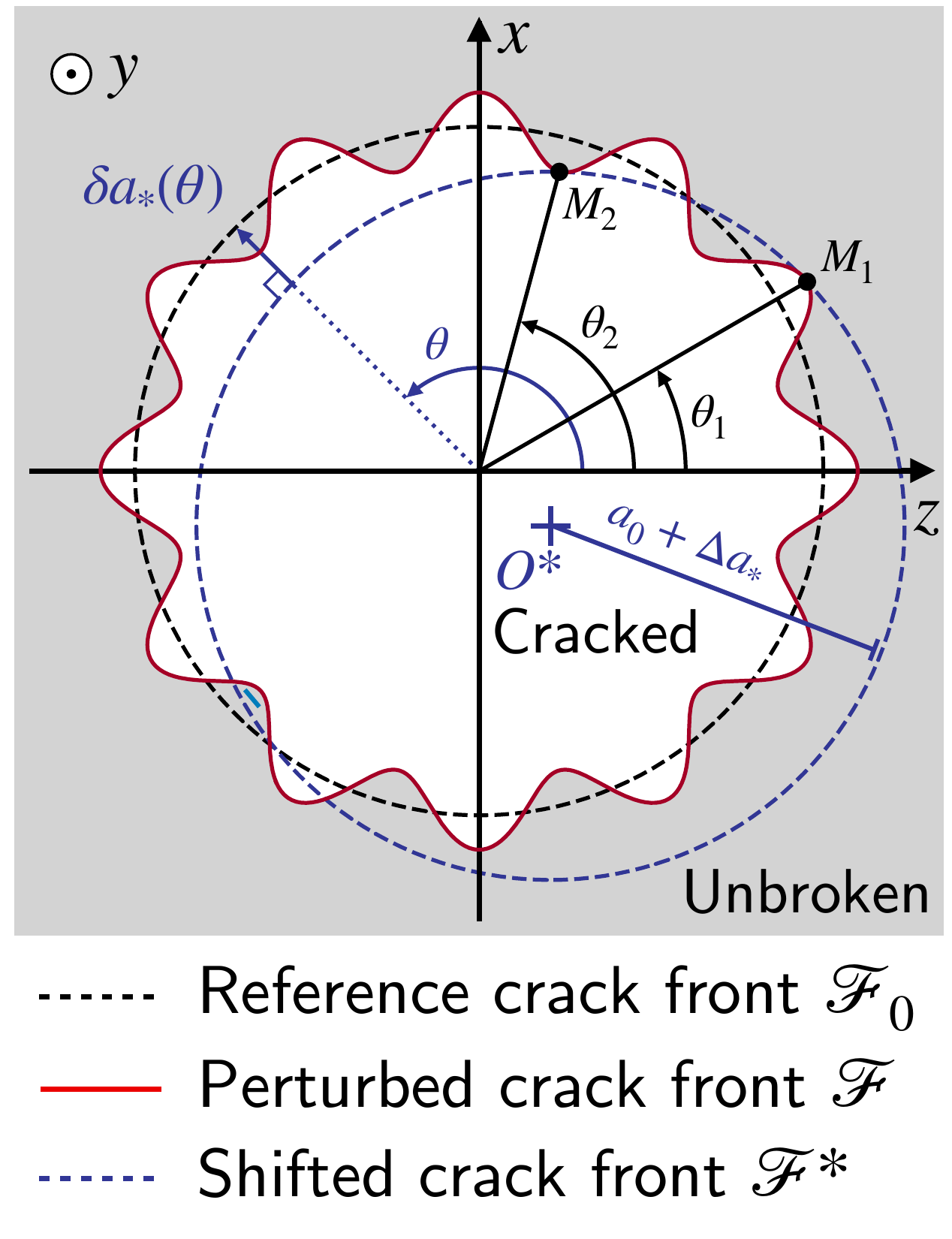}
		\caption{The first-order variations of the kernel $Z^0(a_0; \theta_1, \theta_2)$, associated with the reference circular crack front (in dashed black line), are calculated from a shifted circular crack front $\mathcal{F}_*$ (in dashed blue line) of radius $a_0+\Delta a_*$ and centered in $O^*$, that passes by $M_1$ and $M_2$ located on the perturbed crack front (in red solid line). The transformation from the reference crack front $\mathcal{F}_0$ and shifted crack front $\mathcal{F}^*$ is parametrized by the perturbation $\delta a_*\left(\theta\right)$ defined in Eq.~\eqref{eq:star_transformation_general}.}
		\label{fig:star_transformation}
	\end{figure}

	In the general case where $\delta a$ does not vanish at $M_1$ and $M_2$, one must find a perturbation $\delta a_*$ (i) for which the variations of the fundamental kernel $\delta W^*$ are known \textit{a priori}, and (ii) that takes the same values as $\delta a$ at two points $M_1$ and $M_2$ (see Fig.~\ref{fig:star_transformation}):
	\begin{equation}
		\label{eq:star_transformation_condition}
		\delta a_*(\theta_1) = \delta a(\theta_1) \text{, and } \delta a_*(\theta_2) = \delta a(\theta_2).
	\end{equation}
	One then has \citep{favier_coplanar_2006}:
	\begin{equation}
		\label{eq:perturbated_fundamental_kernel_general}
		\dfrac{\delta W(\theta_1, \theta_2)}{\mathcal{D}(M^0_1, M^0_2)^2} = \dfrac{\delta W^*(\theta_1, \theta_2)}{\mathcal{D}(M^0_1, M^0_2)^2} + \dfrac{1}{2\pi}\mathrm{PV}\int_{0}^{2\pi} \dfrac{W^0(a_0; \theta_1, \theta)}{\mathcal{D}(M^0_1, M^0)^2} \dfrac{W^0(a_0; \theta_2, \theta)}{\mathcal{D}(M^0_2, M^0)^2} \left[\delta a(\theta) - \delta a_*(\theta)\right] a_0 d\theta + \mathcal{O}(A^2),
	\end{equation}
	where $M^0_1$ and $M^0_2$ are the locations defined by the orthogonal projection on $\mathcal{F}_0$ along the direction $\mathbf{e_r}$ of the points $M_1$ and $M_2$, i.e. their position before the application of the transformation $\delta a$ (see Fig.~\ref{fig:perturbed_crack}). The variation of the kernel $Z$ then reads:
	\begin{equation}
		\label{eq:perturbated_kernel_general}
		\begin{aligned}
			\delta Z(\theta_1, \theta_2) & = \dfrac{W^0(a_0; \theta_1, \theta_2) + \delta W(\theta_1, \theta_2)}{2\pi\mathcal{D}(M_1, M_2)^2} - \dfrac{W^0(a_0; \theta_1, \theta_2)}{2\pi\mathcal{D}(M^0_1, M^0_2)^2} \\
			& = Z^0(a_0; \theta_1, \theta_2) \left[\left(\dfrac{\mathcal{D}(M_1, M_2)}{\mathcal{D}(M^0_1, M^0_2)}\right)^{-2}-1\right] + \dfrac{\delta W^*(\theta_1, \theta_2)}{2\pi\mathcal{D}(M^0_1, M^0_2)^2} \\
			& \hspace{3mm} + \mathrm{PV}\int_{0}^{2\pi} Z^0(a_0; \theta_1, \theta) Z^0(a_0; \theta_2, \theta) \left[\delta a(\theta) - \delta a_*(\theta)\right] a_0 d\theta + \mathcal{O}(A^2),
		\end{aligned}
	\end{equation}
	at first order in $A$. In the case of the semi-infinite crack, the first term of $\delta Z$ is of second order in $A$, and reduces thus to zero in the first-order calculations of \cite{leblond_second_2012}. This is not the case for a reference circular crack front, for which it writes as :
	\begin{equation}
		\label{eq:kernel_star_variation}
		Z^0(a_0; \theta_1, \theta_2) \left[\left(\dfrac{\mathcal{D}(M_1, M_2)}{\mathcal{D}(M^0_1, M^0_2)}\right)^{-2}-1\right] = -\dfrac{\delta a(\theta_1) + \delta a(\theta_2)}{8\pi a_0^3\sin^2\left[(\theta_2-\theta_1)/2\right]} + \mathcal{O}(A^2).
	\end{equation}
	The second and third terms of $\delta Z$ in Eq.~\eqref{eq:perturbated_kernel_general} depend on the actual choice of the perturbation $\delta a_*$. To compute the second term analytically, the fundamental kernel of the crack front parametrized by $a^*(\theta) = a_0 + \delta a^*(\theta)$ must be known explicitly. For unbounded solids, \cite{rice_weight_1989} proposed to express $\delta a^*$ as the combination of translations, rotations or scaling, leaving the fundamental kernel unchanged, i.e. $ \delta W^* = 0$. Such a transformation has been used by \cite{leblond_second_2012} to derive the first-order variation of the fundamental kernel $W$ for the semi-infinite coplanar crack loaded in tensile mode I. However, one can show that, in the case investigated here, this transformation contains a rotation so that the perturbation $\delta a_*$ cannot be directly expressed as a normal extension in the direction $\mathbf{e_r}$.
	
	Here, we follow another route and take $\delta a_*$ as the orthogonal projection in the direction $\mathbf{e_r}$, on the reference crack front $\mathcal{F}_0$, of another circular crack front $\mathcal{F}^*$ that goes by $M_1$ and $M_2$ (see Fig.~\ref{fig:star_transformation}). $\mathcal{F}^*$ corresponds to $\mathcal{F}_0$ dilated by a factor $(1+\Delta a_*/a_0)$, and translated by an amount $\Delta r_*$ in the direction $\pi/2+\theta_*$. The perturbed crack front $\mathcal{F}^*$ is thus a circle of radius $a_0 + \Delta a_*$, centered on the point $O^*$ of Cartesian coordinates $(-\Delta r_*\sin\theta_*, \Delta r_*\cos\theta_*)$. The associated perturbation $\delta a_*$ can be expressed as:
	\begin{equation}
		\label{eq:star_transformation_general}
		\delta a_*(\theta) = \Delta r_* \sin(\theta - \theta_*) + \sqrt{(a_0+\Delta a_*)^2 - {\Delta r_*}^2 \cos^2(\theta - \theta_*)} - a_0.
	\end{equation}
	Since the crack front $\mathcal{F}^*$ is circular, its fundamental kernel $W^*$ is also equal to $1$. As a result, $\delta W^* = 0$, and the second term of $\delta Z$ in Eq.~\eqref{eq:perturbated_kernel_general} vanishes.
	
	It leaves us with the calculation of the principal value integral in Eq.~\eqref{eq:perturbated_kernel_general}. There is an infinite number of circles that go by the points $M_1$ and $M_2$. Consequently, $\delta a_*$ is a 3-parameters transformation $(\Delta a_*, \Delta r_*, \theta_*)$ that must satisfy the sole 2 conditions of Eq.~\eqref{eq:star_transformation_condition}. One may then carefully choose an appropriate $\theta_*$ so that (i) $\delta a_*$ of Eq.~\eqref{eq:star_transformation_general} is well-defined, and (ii) one can compute the third term of $\delta Z$ analytically. Here, we take:
	\begin{equation}
		\label{eq:star_transformation_theta_star}
		\theta_* = \dfrac{\theta_1 + \theta_2}{2}.
	\end{equation}
	Note that, in equation~\eqref{eq:perturbated_kernel_general}, one only needs an expression of $\delta a_*$ at first order in $A$ to compute the variations of $Z$. The conditions~\eqref{eq:star_transformation_condition} then yield:
	\begin{equation}
		\label{eq:star_transformation_parameters}
		\Delta r_* = \dfrac{\delta a(\theta_2) - \delta a(\theta_1)}{2\sin\left[(\theta_2-\theta_1)/2\right]} + \mathcal{O}(A^2),
		\text{and } \Delta a_* = \dfrac{\delta a(\theta_1) + \delta a(\theta_2)}{2} + \mathcal{O}(A^2).
	\end{equation}
	The perturbation $\delta a_*$ can finally be expressed as:
	\begin{equation}
		\label{eq:star_transformation_1stOrder}
		\begin{aligned}
			& \delta a_*(\theta) = \dfrac{\delta a(\theta_1) + \delta a(\theta_2)}{2} + \dfrac{\delta a(\theta_2) - \delta a(\theta_1)}{2} \cdot \dfrac{\sin\left[\theta - (\theta_1+\theta_2)/2\right]}{\sin\left[(\theta_2-\theta_1)/2\right]} + \mathcal{O}(A^2) \\
			& \hspace{1mm} = \delta a(\theta_1) + \dfrac{\delta a(\theta_2) - \delta a(\theta_1)}{2} \left[\dfrac{\cos\left[(\theta_2-\theta_1)/2\right]}{\sin\left[(\theta_2-\theta_1)/2\right]} \sin(\theta-\theta_1) + 2\sin^2[(\theta-\theta_1)/2]\right] + \mathcal{O}(A^2) \\
			& \hspace{1mm} = \delta a(\theta_2) + \dfrac{\delta a(\theta_2) - \delta a(\theta_1)}{2} \left[\dfrac{\cos\left[(\theta_2-\theta_1)/2\right]}{\sin\left[(\theta_2-\theta_1)/2\right]} \sin(\theta-\theta_2) - 2\sin^2[(\theta-\theta_2)/2]\right] + \mathcal{O}(A^2).
		\end{aligned}
	\end{equation}
	The second and third expressions of $\delta a_*(\theta)$ in Eq.~\eqref{eq:star_transformation_1stOrder} only involves $(\theta-\theta_1)$ and $(\theta-\theta_2)$ respectively, thanks to the careful choice of $\theta_*$. They are analogous to \cite{leblond_second_2012}'s Eq.~(7), and prove key in the evaluation of the principal value integral of our Eq.~\eqref{eq:perturbated_kernel_general} (see \ref{apdx:calculation_Z} for details).
	
	Next, we use another trick, analogous to the partial fraction decomposition of \cite{leblond_second_2012}'s Eq.~(9), and propose the following decomposition:
	\begin{equation}
		\label{eq:partial_fraction_decomposition}
		\begin{aligned}
			\dfrac{1}{\sin^2[(\theta-\theta_1)/2]\sin^2[(\theta-\theta_2)/2]} = & \dfrac{1}{\sin^2\left[(\theta_2-\theta_1)/2\right]} \left[\dfrac{\cos^2[(\theta-\theta_1)/2]}{\sin^2[(\theta-\theta_1)/2]} + \dfrac{\cos^2[(\theta-\theta_2)/2]}{\sin^2[(\theta-\theta_2)/2]}\right] \\
			& + \dfrac{2\cos[(\theta_2-\theta_1)/2]}{\sin^3[(\theta_2-\theta_1)/2]} \left[ \dfrac{\cos^3[(\theta-\theta_1)/2]}{\sin[(\theta-\theta_1)/2]} - \dfrac{\cos^3[(\theta-\theta_2)/2]}{\sin[(\theta-\theta_2)/2]} \right] \\
			& + \dfrac{2}{\sin^2\left[(\theta_2-\theta_1)/2\right]} \left[\cos^2[(\theta-\theta_1)/2] + \cos^2[(\theta-\theta_2)/2]\right]
		\end{aligned}
	\end{equation}
	Combining Eqs.~\eqref{eq:perturbated_kernel_general}, (\ref{eq:star_transformation_1stOrder}-\ref{eq:partial_fraction_decomposition}), and after some simplifications, one gets:
	\begin{equation}
		\label{eq:perturbated_kernel_1stOrder}
		\scalemath{0.95}{
			\begin{aligned}
				\delta Z~(\theta_1, \theta_2) & = -\dfrac{5}{32\pi a_0^3} \dfrac{\delta a(\theta_1) + \delta a(\theta_2)}{\sin^2\left[(\theta_2-\theta_1)/2\right]} +  \dfrac{1}{(8\pi)^2 a_0^3\sin^2\left[(\theta_2-\theta_1)/2\right]} \int_{0}^{2\pi} 2\left[\cos^2\left(\dfrac{\theta-\theta_1}{2}\right) + \cos^2\left(\dfrac{\theta-\theta_2}{2}\right)\right] \delta a(\theta) d\theta \\
				& + \dfrac{1}{(8\pi)^2 a_0^3\sin^2\left[(\theta_2-\theta_1)/2\right]} \times \Bigg\lbrace \\
				& \hspace{10mm} \left. \mathrm{PV}\int_{0}^{2\pi} \dfrac{\cos^2[(\theta-\theta_1)/2]}{\sin^2[(\theta-\theta_1)/2]} \left[\delta a(\theta) - \delta a(\theta_1)\right] d\theta
				+ \mathrm{PV}\int_{0}^{2\pi} \dfrac{\cos^2[(\theta-\theta_2)/2]}{\sin^2[(\theta-\theta_2)/2]} \left[\delta a(\theta) - \delta a(\theta_2)\right] d\theta \right. \\
				& \hspace{10mm} \left. + \dfrac{2\cos[(\theta_2-\theta_1)/2]}{\sin^3[(\theta_2-\theta_1)/2]} \cdot \mathrm{PV}\int_{0}^{2\pi} \left[ \dfrac{\cos^3[(\theta-\theta_1)/2]}{\sin[(\theta-\theta_1)/2]} - \dfrac{\cos^3[(\theta-\theta_2)/2]}{\sin[(\theta-\theta_2)/2]} \right] \delta a(\theta) d\theta \right\rbrace + \mathcal{O}(A^2)
			\end{aligned}
		}
	\end{equation}
	
	Details on the derivation of Eq.~\eqref{eq:perturbated_kernel_1stOrder} are given in \ref{apdx:calculation_Z}. Note that the two first terms of Eq.~\eqref{eq:perturbated_kernel_1stOrder} typically emerge from the finiteness of the crack geometry, and go to zero in the limit $a_0 \rightarrow +\infty$. One can show that \cite{leblond_second_2012}'s Eq.~(10) for the semi-infinite crack is retrieved in the same limit.
	
	We derived here the first-order variation of the kernel $Z$ for a penny-shaped crack loaded in mode I. However, the determination of $\delta a_*$ provided here enables extending our results to mixed mode I+II+III \citep{favier_coplanar_2006}, to other geometries like e.g. the circular connection \citep{gao_nearly_1987}, or to circular cracks propagating in cohesive materials \citep{lebihain_cohesive_2022}.
	
	\subsection{Second-order expansion of the SIF}
	
	We consider again a crack front $\mathcal{F}$ that has been perturbed from a circular configuration of radius $a_0$ by an amount $\delta a(\theta) = A\phi(\theta)$, where $A$ is now a \emph{small} albeit \emph{not infinitesimal} parameter. The stress intensity factor $\KI$ and the kernel $Z$ along the perturbed crack front $\mathcal{F}$ can be expanded at several orders in $A$. Their expansion \review{writes}:
	\begin{equation}
	\modif{
		\label{eq:SIF_2ndOrder_general}
		\begin{cases}
			\begin{aligned}
				\KI(a_0, \left[A\phi\right]; \theta_1) = \KI^0(a_0, \theta_1) & + A\,\KI^1(a_0, [\KI^0], [\phi]; \theta_1) \\
				&	+ A^2\KI^2(a_0, [\KI^0], [\phi]; \theta_1) + \mathcal{O}(A^3)
			\end{aligned}\\
			Z(a_0, \left[A\phi\right]; \theta_1, \theta_2) = Z^0(a_0; \theta_1, \theta_2) + AZ^1(a_0, \left[\phi\right]; \theta_1, \theta_2) + \mathcal{O}(A^2),
		\end{cases}
	}
	\end{equation}
	where $\KI^1(a_0, [\KI^0], [\phi]; \theta_1)$ is the first-order variation of the stress intensity factor induced at $\theta_1$ by the perturbation $\phi$ applied to a reference circular configuration of radius $a_0$ and SIF $\KI^0$, and $\KI^2(a_0, [\KI^0], [\phi]; \theta_1)$ its second-order variation. Similarly, $Z^1(a_0, \left[\phi\right]; \theta_1, \theta_2)$ is the first-order variation of the kernel $Z^0$ induced at $(\theta_1, \theta_2)$ by the perturbation $\phi$ from the reference configuration of radius $a_0$. $\KI^1$ and $Z^0$ have been derived by \cite{gao_somewhat_1987}, and are recalled in Eqs.~(\ref{eq:perturbed_SIF_1stOrder}-\ref{eq:fundamental_kernel}). $Z^1$ has been derived here in Eq.~\eqref{eq:perturbated_kernel_1stOrder}. Our goal here is to provide an analytical expression for $\KI^2$. 
	
	To this end, we follow the steps of \cite{leblond_second_2012}, and superimpose to the primary perturbation $\delta a$ a \emph{proportional} yet \emph{infinitesimal} secondary perturbation $\delta A \phi$, where $\delta A$ is an infinitesimally small parameter. By definition, one has:
	\begin{equation}
		\label{eq:SIF_variation_01}
		\begin{aligned}
			\KI(a_0, \left[(A+\delta A)\phi\right]; \theta_1) - \KI(a_0, \left[A\phi\right]; \theta_1) & = \dfrac{\partial \KI}{\partial A}(a_0, \left[A\phi\right]; \theta_1) \delta A + \mathcal{O}(\delta A^2) \\
			& = \left\lbrace\KI^1(a_0, [\KI^0], [\phi]; \theta_1) + 2A\KI^2(a_0, [\KI^0], [\phi]; \theta_1) + \mathcal{O}(A^2)\right\rbrace \delta A + \mathcal{O}(\delta A^2)
		\end{aligned}
	\end{equation}
	\modif{Alternatively, one may consider the front $\mathcal{F'}$ of reference radius $a'_0 = a_0 + \delta A \phi(\theta_1)$ perturbed by the quantity $\delta a(\theta) = A\phi(\theta)$, and perturbed it further by a infinitesimal quantity $\delta A \phi(\theta) - \delta A \phi(\theta_1)$. Yet, this secondary perturbation is not performed in the direction $\mathbf{e_r}$, but rather in the direction normal to the perturbed front $\mathcal{F'}$. Given that the error in the resulting front position is of second order in both in $\delta A$ (see \cite{leblond_second_2012})\footnote{\modif{Note that one must ensure that the secondary perturbation is zero at $\theta = \theta_1$, otherwise the transformation induces a phase-shift that is first-order in $\delta A$. This was overlooked by \cite{leblond_second_2012} and \cite{vasoya_second_2013}, but their results remain valid as these first-order corrections are zero when $\KI^0$ is independent of the position along the front.}}, one has \citep{rice_weight_1989}:}
	\begin{equation}
		\label{eq:SIF_variation_02}
		\modif{
		\begin{aligned}
			& \KI(a_0, \left[(A+\delta A)\phi\right]; \theta_1) - \KI(a_0, \left[A\phi\right]; \theta_1) = \KI(a_0+\delta A\phi(\theta_1), \left[A\phi\right]; a_0; \theta_1) \\
			& \hspace{2cm} + \mathrm{PV}\int_{0}^{2\pi} Z(a_0+\delta A\phi(\theta_1), \left[A\phi\right]; \theta_1, \theta) \KI(a_0+\delta A\phi(\theta_1), \left[A\phi\right];\theta) \left[\delta A\phi(\theta) - \delta A\phi(\theta_1)\right] ds(\theta) \\
			& \hspace{2cm} - \KI(a_0, \left[A\phi\right]; \theta_1) + \mathcal{O}(\delta A^2),
		\end{aligned}
		}
	\end{equation}
	where $s$ is the curvilinear abscissa along the perturbed front $\mathcal{F'}$. Equating Eqs.~\eqref{eq:SIF_variation_01} and \eqref{eq:SIF_variation_02}, one may identify the terms of zero order in $A$ in the expression of $\partial \KI/\partial A$:
	\begin{equation}
		\label{eq:perturbed_SIF_K1}
		\KI^1(a_0, [\KI^0], [\phi]; \theta_1) = \dfrac{\partial \KI^0}{\partial a}(a_0, \theta_1) \phi(\theta_1) + \mathrm{PV}\int_{0}^{2\pi} Z^0(a_0; \theta_1, \theta) \KI^0(a_0, \theta) \left[\phi(\theta) - \phi(\theta_1)\right] a_0 d\theta,
	\end{equation}
	which is Eq.~\eqref{eq:perturbed_SIF_1stOrder}. Identifying the terms of first order in $A$ in the expression of $\partial \KI/\partial A$, one has after some simplifications:
	\begin{equation}
		\label{eq:perturbed_SIF_K2}
		\begin{aligned}
			\KI^2(a_0, [\KI^0], [\phi]; \theta_1) = & \frac{1}{2}\dfrac{\partial^2 \KI^0}{\partial a}(a_0, \theta_1) \phi(\theta_1)^2 + \frac{1}{2}\mathrm{PV}\int_{0}^{2\pi} Z^0(a_0; \theta_1, \theta) \dfrac{\partial\KI^0}{\partial a}(a_0, \theta) \left[\phi(\theta)^2 - \phi(\theta_1)^2\right] a_0 d\theta \\
			& + \frac{1}{2}\mathrm{PV}\int_{0}^{2\pi} Z^0(a_0; \theta_1, \theta) \KI^0(a_0, \theta) \left[\phi(\theta) - \phi(\theta_1)\right]^2 a_0 d\theta \\
			& + \frac{1}{2}\mathrm{PV}\int_{0}^{2\pi} Z^1(a_0, \phi; \theta_1, \theta) \KI^0(a_0, \theta) \left[\phi(\theta) - \phi(\theta_1)\right] a_0 d\theta\\
			& + \frac{1}{2}\mathrm{PV}\int_{0}^{2\pi}\int_{0}^{2\pi} Z^0(a_0; \theta_1, \theta) Z^0(a_0; \theta, \theta') \KI^0(a_0, \theta') \left[\phi(\theta) - \phi(\theta_1)\right] \left[\phi(\theta') - \phi(\theta)\right] a_0^2 d\theta d\theta'.
		\end{aligned}
	\end{equation}
	Details on the derivation of Eq.~\eqref{eq:perturbed_SIF_K2} can be found in \ref{apdx:calculation_K}. Equations~\eqref{eq:perturbed_SIF_K1} and \eqref{eq:perturbed_SIF_K2} are general equations that prove unfit for analytical calculations, and one has to resort to Fourier series to evaluate them. We shall do it here in the simplified case where the crack is loaded through an axisymmetric system of forces, i.e. when $\KI^0$ is independent of $\theta$. Note however that this is enough to derive second-order analytical solutions for crack propagation in heterogeneous media, even when $\KI^0$ actually depends on $\theta$, as these fluctuations are already of second order in the first-order theory of \cite{gao_somewhat_1987}. However, for the sake of completeness, the derivation for a non-axisymmetric $\KI^0$ is given in \ref{apdx:calculation_K}.
	
	Next, we assume that $\KI^0$ only depends on the radius $a_0$ of the reference penny-shaped crack. One can decompose $\phi$ in Fourier series. It reads:
	\begin{equation}
		\label{eq:modal_perturbation}
		\phi(\theta) = \sum_{k=-\infty}^{+\infty} \hat{\phi}_k e^{ik\theta} \text{, where } \hat{\phi}_k = \dfrac{1}{2\pi} \int_{0}^{2\pi} \phi(\theta) e^{-ik\theta} d\theta.
	\end{equation}
	Then, the first-order contribution of Eq.~\eqref{eq:perturbed_SIF_K1} writes as:
	\begin{equation}
		\label{eq:fourier_SIF_K1}
		\KI^1(a_0, [\KI^0], [\phi]; \theta_1) = \sum_{k=-\infty}^{+\infty} \left[\dfrac{d\KI^0}{da}(a_0) - \dfrac{\abs{k}}{2a_0}\KI^0(a_0)\right]\hat{\phi}_k e^{ik\theta_1},
	\end{equation}
	and the second-order contribution of Eq.~\eqref{eq:perturbed_SIF_K2} reads:
	\begin{equation}
		\label{eq:fourier_SIF_K2}
		\scalemath{0.95}{
			\KI^2(a_0, [\KI^0], [\phi]; \theta_1) = \sum_{k=-\infty}^{+\infty} \sum_{k'=-\infty}^{+\infty} \left[\dfrac{1}{2}\dfrac{d^2\KI^0}{da^2}(a_0) - \dfrac{1}{4a_0} \dfrac{d\KI^0}{da}(a_0) \abs{k+k'} + \dfrac{R(k,k')}{16a_0^2}\KI^0(a_0) + \dfrac{\abs{k}+\abs{k'}}{4a_0^2}\KI^0(a_0)\right] \hat{\phi}_k\hat{\phi}_{k'} e^{i(k+k')\theta_1},
		}	
	\end{equation}
	where $R(k,k') = 2\abs{k+k'}(\abs{k}+\abs{k'}) -(k+k')^2 -k^2 - {k'}^2$ is twice the function $F$ of \cite{leblond_second_2012}, written in the form simplified by \cite{vasoya_experimental_2016}. Note that Eq.~\eqref{eq:perturbed_SIF_K2} differs from Eq.~(8) of \cite{vasoya_second_2013} only by the term $(\abs{k}+\abs{k'})\KI^0(a_0)/4a_0^2$ that typically emerges from the finiteness of the penny-shaped crack geometry. As expected, this term goes to $0$ when $a_0 \rightarrow +\infty$, and one retrieves, in this limit, the results of \cite{leblond_second_2012} and \cite{vasoya_second_2013} for the semi-infinite crack.\\
	
	One can alternatively express Eq.~(\ref{eq:SIF_2ndOrder_general}a) in terms of the Fourier coefficient $\widehat{\delta a}_k$ of the perturbation $\delta a$:
	\begin{equation}
		\label{eq:fourier_SIF_2ndOrder}
		\KI(a_0, \left[\delta a\right]; \theta_1) = \KI^0(a_0) \left[ 1 + \sum_{k} L(k) \dfrac{\widehat{\delta a}_k}{a_0} e^{ik\theta_1} + \sum_{k,k'} H(k, k') \dfrac{\widehat{\delta a}_k}{a_0} \dfrac{\widehat{\delta a}_{k'}}{a_0} e^{i(k+k')\theta_1} \right] + \mathcal{O}(A^3),
	\end{equation}
	where the first- and second-order convolution kernels $L$ and $H$ read:
	\begin{equation}
		\label{eq:fourier_kernels_2ndOrder}
		\begin{cases}
			L(k) = \dfrac{a_0}{\KI^0(a_0)} \dfrac{d\KI^0}{da}(a_0) - \dfrac{\abs{k}}{2} \\
			H(k,k') = \dfrac{a_0^2}{2\KI^0(a_0)} \dfrac{d^2\KI^0}{da^2}(a_0) - \dfrac{a_0}{\KI^0(a_0)} \dfrac{d\KI^0}{da}(a_0) \dfrac{\abs{k+k'}}{4} + \dfrac{R(k,k')}{16} + \dfrac{\abs{k}+\abs{k'}}{4}.
		\end{cases}
	\end{equation}
	
	\section{Validation of the proposed theory}
	\label{sec:validation}
	
	We aim now at validating the second-order expansion of the stress intensity factor given in Eq.~\eqref{eq:fourier_SIF_2ndOrder}. One may consider that retrieving \cite{leblond_second_2012}'s and \cite{vasoya_second_2013}'s results in the limit of very large crack radius $a_0$ proves enough to validate our calculations. Yet, \cite{leblond_second_2012}'s second-order formula has been found conflicting with those previously derived by \cite{adda-bedia_second-order_2006} and \cite{katzav_roughness_2007}, and its validity has been checked through \emph{numerical simulations} only. As a consequence, one should seek further validation cases to assess the correctness of Eq.~\eqref{eq:fourier_SIF_2ndOrder}. Here, we follow two different routes: first, (i) we consider elliptical cracks as perturbed circular cracks and test our formula against \emph{analytical results} obtained for this geometry \citep{irwin_fracture_1958}. It provides a critical test for our second-order theory, and yields a \emph{first analytical validation} of \cite{leblond_second_2012}'s formula for the semi-infinite crack. Next, (ii) we use \cite{lazarus_numerics_2003}'s model that computes the mode I SIF variations at all orders to compare our second-order expansions to numerical results obtained for circular cracks deformed by cosinusoidal perturbations of varying modes and amplitudes. This ultimately allows us to estimate the validity range of our second-order theory.
	
	\subsection{SIF distribution along the front of an elliptical crack}
	
	In contrast with models based on the semi-infinite crack geometry \citep{leblond_second_2012, vasoya_second_2013}, our asymptotic theory derived for penny-shaped cracks can be tested against analytical solutions obtained for planar cracks with elliptical front shapes. Here, we consider a crack whose front describes an ellipsis centered in $O$, of short semi-axis $c$ aligned with $\mathbf{e_z}$, and long semi-axis $b$ aligned with $\mathbf{e_x}$. In the polar coordinate system of Fig.~\ref{fig:perturbed_crack}, the front position is parametrized by the function:
	\begin{equation}
		\label{eq:ellipsis_transformation}
		a(\theta) = \dfrac{bc}{\sqrt{b^2\sin^2(\theta) + c^2\cos^2(\theta)}}.
	\end{equation}
	The crack is loaded by a uniform remote tensile stress $\sigma$. The distribution of mode I stress intensity factor along the elliptical crack front is known \emph{exactly} from the work of \cite{irwin_fracture_1958}. It reads:
	\begin{equation}
		\label{eq:ellipsis_KI_exact}
		\KI^e(\theta) = \dfrac{\sigma\sqrt{\pi c}}{\mathrm{E}(\kappa)} \left(\dfrac{c^4/b^4 \cos^2(\theta) + \sin^2(\theta)}{c^2/b^2 \cos^2(\theta) + \sin^2(\theta)}\right)^{1/4},
	\end{equation}
	where $\mathrm{E}(\kappa)$ is the elliptical integral of the second kind evaluated at $\kappa = \left(1-c^2/b^2\right)^{1/2}$. When $b = c = a_0$, one retrieves the case of a circular crack front of radius $a_0$:
	\begin{equation}
		\label{eq:ellipsis_KI_reference}
		\KI^0(a_0) = \dfrac{2}{\pi} \sigma \sqrt{\pi a_0},
	\end{equation}
	so that the convolution kernels of Eq.~\eqref{eq:fourier_kernels_2ndOrder} read:
	\begin{equation}
		\label{eq:ellipsis_fourier_kernels}
		\begin{cases}
			L(k) = - \dfrac{\abs{k}}{2} + \dfrac{1}{2} \\ \vspace{2mm}
			H(k,k') = \dfrac{R(k,k')}{16} + \dfrac{\abs{k}+\abs{k'}}{4} - \dfrac{\abs{k+k'}}{8} - \dfrac{1}{8}.
		\end{cases}
	\end{equation}
	
	To derive asymptotic values of $\KI^e$ from our second-order theory, one needs to choose (i) a small parameter $\epsilon$ that quantifies the deviation from circularity of the elliptical crack front, and (ii) the radius $a_0$ of the reference circular configuration. We follow here the reasoning of \cite{gao_somewhat_1987}, and take $\epsilon = 1-c/b$. However, they considered a reference radius $a_0$ equal to the crack front position at the point of SIF evaluation, i.e. $a_0 = a(\theta_1)$. Consequently, the perturbation $\delta a(\theta) = a(\theta) - a_0$ is zero at $\theta = \theta_1$, and the double sum of Eq.~\eqref{eq:fourier_SIF_K2} containing terms in $\abs{k}+\abs{k'}$:
	\begin{equation}
		\sum_{k,k'} \dfrac{\abs{k}+\abs{k'}}{4} \widehat{\delta a}_{k} \widehat{\delta a}_{k'} e^{i(k+k')\theta_1} = \left[\sum_{k}  \dfrac{\abs{k}}{2} \widehat{\delta a}_k e^{i(k+k')\theta_1}\right] \times \delta a(\theta_1) 
	\end{equation}
	is also null. Yet, these terms are precisely those we need to check first, as they vanish in the limit of very large crack radii, and cannot be tested against \cite{vasoya_second_2013}'s formula for the semi-infinite crack. Here, we rather take the length $c$ of the short semi-axis as reference radius $a_0$. Following Eq.~\eqref{eq:ellipsis_transformation}, the perturbation $\delta a = a - a_0$ can be expanded as:
	\begin{equation}
		\label{eq:ellipsis_transformation_expansion}
		\begin{aligned}
			\delta a(\theta)/a_0 =\, & \epsilon A_1(\theta) + \epsilon^2 A_2(\theta) + \mathcal{O}(\epsilon^3) \\
			=\, & \epsilon \left[\frac{1}{2}+\frac{1}{2}\cos(2\theta)\right] + \epsilon^2 \left[\frac{5}{16}+\frac{1}{2}\cos(2\theta)+\frac{3}{16}\cos(4\theta)\right] + \mathcal{O}(\epsilon^3),
		\end{aligned}
	\end{equation}
	where $A_1$ and $A_2$ are the first- and second-order expansions of $\delta a/a_0$. Similarly, the exact SIF distribution $\KI^e$ given in Eq.~\eqref{eq:ellipsis_KI_exact} can be expanded to second-order in $\epsilon$:
	\begin{equation}
		\label{eq:ellipsis_KI_exact_expansion}
		\KI^e(\theta) / \KI^0\left(a_0\right) = 1 + \epsilon\left[\frac{1}{4}-\frac{1}{4}\cos(2\theta)\right] + \epsilon^2\left[\frac{11}{64}-\frac{3}{16}\cos(2\theta)-\frac{11}{64}\cos(4\theta)\right] + \mathcal{O}(\epsilon^3).
	\end{equation}
	One seeks now to find back Eq.~\eqref{eq:ellipsis_KI_exact_expansion} from our non-linear theory. Using  Eq.~\eqref{eq:ellipsis_KI_exact_expansion} in Eq.~\eqref{eq:fourier_SIF_2ndOrder}, the second-order estimate $\KI^p$ of the mode I SIF reads:
	\begin{equation}
		\label{eq:ellipsis_2ndOrder}
		\KI^p(\theta)/ \KI^0(a_0) = 1 + \epsilon \left[\sum_{k} L(k) \hat{A}^1_k e^{ik\theta}\right] + \epsilon^2 \left[\sum_{k} L(k) \hat{A}^2_k e^{ik\theta} + \sum_{k,k'} H(k,k') \hat{A}^1_k \hat{A}^1_{k'} e^{i(k+k')\theta}\right] + \mathcal{O}(\epsilon^3),
	\end{equation}
	where $\hat{A}^1_k$ and $\hat{A}^2_k$ are the Fourier coefficients of $A_1$ and $A_2$. Using Eqs.~\eqref{eq:ellipsis_fourier_kernels} and \eqref{eq:ellipsis_transformation_expansion}, one finds:
	\begin{equation}
		\label{eq:ellipsis_L_Phi1}
		\begin{aligned}
			& \sum_{k} L(k) \hat{A}^1_k e^{ik\theta} = \dfrac{1}{4} - \dfrac{1}{4}\cos(2\theta), \\[8pt]
			& \sum_{k} L(k) \hat{A}^2_k e^{ik\theta} = \dfrac{5}{32} - \dfrac{1}{4}\cos(2\theta) - \dfrac{9}{32}\cos(4\theta), \\[8pt]
			\text{and } & \sum_{k,k'} H(k,k') \hat{A}^1_k \hat{A}^1_{k'} e^{i(k+k')\theta} = \dfrac{1}{64} + \dfrac{1}{16}\cos(2\theta) + \dfrac{7}{64}\cos(4\theta),
		\end{aligned}
	\end{equation}
	which yields the result expected from Eq.~\eqref{eq:ellipsis_KI_exact_expansion}. Note that we also checked that $\KI^e$ and our expansion $\KI^p$ are equal up to second-order when one takes $a_0 = a(\theta_1)$ as \cite{gao_somewhat_1987} did. Further validations of our second-order theory are provided in \ref{apdx:validation_K}, even for the case of a non-axisymmetric $\KI^0$.\\
	
	The correctness of our Eq.~\eqref{eq:fourier_SIF_2ndOrder} in turn provides a first analytical validation of \cite{leblond_second_2012}'s formula, which is obtained in the limit $a_0 \rightarrow +\infty$. Our non-linear perturbative approach yields estimates of the stress intensity factor along elliptical crack fronts for $b/c \simeq 1$, even when it arises from discontinuous stress distributions (e.g. a point-force loading). It is complementary to other methods available in the literature that provide $\KI$ estimates along elliptical crack fronts for arbitrary $b/c$ ratio, but considering polynomial stress distributions \citep{shah_stress_1971, kassir_3D_1975, atroshchenko_stress_2009}. However, as we will see next, its scope is much wider as it can be used on perturbed fronts of arbitrary shapes.
	
	\subsection{SIF distribution along a sinusoidal crack front}
	
	Now that our second-order theory has been validated considering elliptical crack fronts, one needs to assess its performances in estimating the stress intensity factor distribution along randomly perturbed crack fronts. As elliptical crack fronts are associated with second-order perturbations consisting of the superposition of several deformation modes (see Eq.~\eqref{eq:ellipsis_transformation_expansion}), they do not allow us to quantify the errors made by our second-order model for a given perturbation mode of fixed amplitude. We rather investigate a wavy crack front described by:
	\begin{equation}
		\label{eq:cosinus_perturbation}
		a(\theta) = a_0 \left[1 + \epsilon \cos(k\theta)\right],
	\end{equation}
	where the integer $k \geq 2$ is the mode of the perturbation, and $\epsilon$ is a small parameter that quantifies its amplitude. We assume next that the crack parametrized by Eq.~\eqref{eq:cosinus_perturbation} is loaded by a uniform remote tensile stress $\sigma$. As there is no analytical formula available for the mode I stress intensity factor along the front of such wavy cracks, one needs to compute $\KI$ numerically. Here, we use the method of \cite{lazarus_numerics_2003} and \cite{david_key_2022} that computes $\KI$ along an arbitrary crack front using the perturbative approach of \cite{rice_weight_1989} iteratively. Namely, (i) we start from a penny-shaped configuration for which the mode I SIF $\KI$ and the fundamental kernel $W$ are known. Then, (ii) we discretize the crack front in $N$ equidistant points, and sequentially deform it by a small amount $\delta a(\theta) = \Delta \epsilon \cos(k\theta)$ with $\Delta\epsilon = 0.0025$, and (iii) update both $\KI$ and $W$ using formulae similar to Eqs.~\eqref{eq:perturbed_SIF_1stOrder} and \eqref{eq:perturbated_fundamental_kernel_raw}. As $\Delta\epsilon \ll 1$, the first-order assumption of \cite{rice_weight_1989} applies, and $\KI$ is computed at \emph{all orders} in $\epsilon$. The update of the stress intensity factor and the fundamental kernel are costly procedures, but this method still demands less computational resources than e.g. simulations based on the finite element method (FEM) or the boundary element method (BEM), as only the crack front needs to be meshed. This method is also powerful in modeling far-field loading applied at infinity or hypotheses of infinite body.
	
	Using our second-order model of Eq.~\eqref{eq:fourier_SIF_2ndOrder}, together with the convolution kernels of Eq.~\eqref{eq:ellipsis_fourier_kernels}, the stress intensity factor variations $\delta \KI^p$ from the uniform $\KI^0$ reference reads:
	\begin{equation}
		\label{eq:cosinus_SIF_variations}
		\delta \KI^p(\theta)/\KI^0(a_0) = - \dfrac{k-1}{2} \epsilon \cos(k\theta) - \dfrac{k^2 - 4k + 1}{16} \epsilon^2 + \dfrac{k^2 + 2k - 1}{16} \epsilon^2 \cos(2k\theta) + \mathcal{O}(\epsilon^3),
	\end{equation}
	where $\KI^0(a_0)$ is defined in Eq.~\eqref{eq:ellipsis_KI_reference}.\\
	
	\begin{figure*}
		\centering
		\noindent\includegraphics[width=0.85\textwidth]{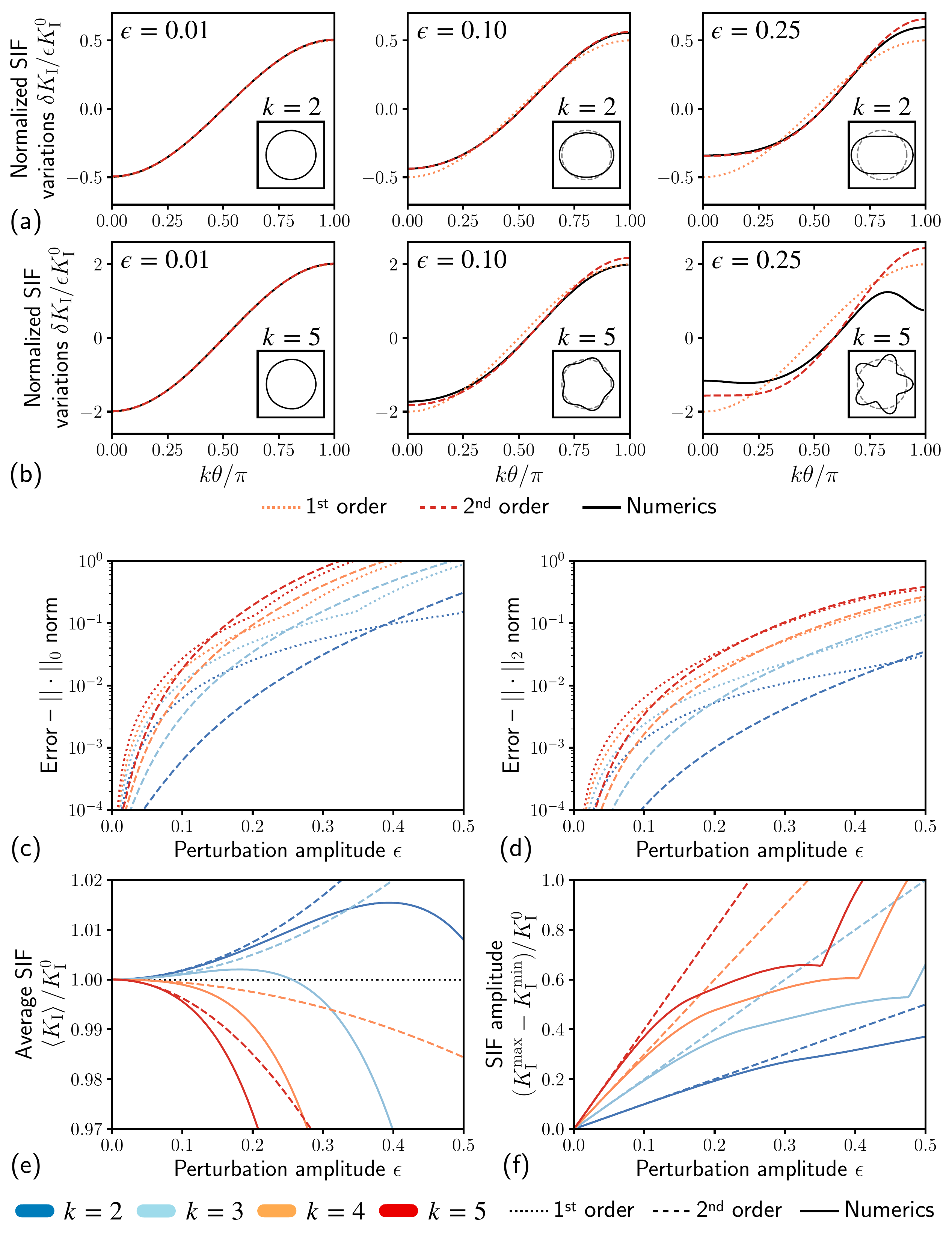}
		\caption{Normalized stress intensity factor variations $\delta \KI/\epsilon\KI^0$ along a crack front perturbed by $\delta a(\theta)/a_0 = \epsilon \cos(k\theta)$ for (a) $k=2$ and (b) $k=5$, and $\epsilon = 0.01$ (left panel), $\epsilon = 0.1$ (middle panel), and $\epsilon = 0.25$ (right panel). The SIF variations $\KI^\mathrm{num}$ computed from \cite{lazarus_numerics_2003}'s numerical model (in black solid line) are compared to the first-order estimate of \cite{gao_somewhat_1987} (in dotted orange line) and our second-order prediction (in dashed red line). The errors on the SIF variations between the numerical simulations (in solid line) and the first-order (in dotted line) and second-order (in dashed line) models are compared using (c) $L^0$ (maximum) or (d) $L^2$ (Euclidean) norms. They increase with both the perturbation mode $k$ and amplitude $\epsilon$. Deviations of the numerical results from the first- and second-order predictions can be directly observed on (e) the average $\left<\KI\right>$ of the stress-intensity factor along the crack front, or (f) its amplitude $\KI^\mathrm{max}-\KI^\mathrm{min}$, for which first and second-order estimates coincide.}
		\label{fig:validation_cosinus}
	\end{figure*}
	
	The numerical results using the method of \cite{lazarus_numerics_2003} (in black solid line) are compared to the second-order prediction of Eq.~\eqref{eq:cosinus_SIF_variations} (in dashed red line) and to its first-order truncation (in dotted orange line) in Fig.~\ref{fig:validation_cosinus}. For $k=2$ and $\epsilon = 0.01$ (see Fig.~\ref{fig:validation_cosinus}a), the three quantities are close. When $\epsilon=0.1$, the first-order assumption is no longer valid, and the linear estimate departs from the numerical results, while the second-order prediction remains accurate. As $\epsilon$ grows even larger ($\epsilon=0.25$), the second-order prediction starts overestimating the stress intensity factor on the regions where the crack is less elongated. Note however that the second-order model do predict an upward shift in the average of the SIF distribution, while the first-order prediction remains symmetric with respect to $\delta\KI=0$. For $k=5$ (see Fig.~\ref{fig:validation_cosinus}b), the validity range of our theory shrinks, and departure of the second-order prediction from the numerical results is already noticeable when $\epsilon = 0.1$. When $\epsilon=0.25$, the discrepancies are fairly large, and our model fails in predicting the strong decrease of SIF amplitude at $\theta = \pi/5$. We observe that the errors made by our second-order model around this point are even larger than those made by the first-order theory. These preliminary observations are further investigated in Fig.~\ref{fig:validation_cosinus}c-d where we show the evolution with $\epsilon$ of the (maximal) error using the $L^0$ norm:
	\begin{equation*}
		\vert\vert\, \mathrm{error} \,\vert\vert_0 = \max_{\theta \in \left[0, 2\pi\right]} \abs{\delta\KI^p(\theta) - \delta\KI^\mathrm{num}(\theta)}/\KI^0(a_0),
	\end{equation*}
	and the (average) error using the $L^2$ norm:
	\begin{equation*}
		\vert\vert\, \mathrm{error} \,\vert\vert_2 = \dfrac{1}{2\pi /\KI^0(a_0)} \int_{0}^{2\pi} \left(\delta\KI^p(\theta) - \delta\KI^\mathrm{num}(\theta)\right)^2 d\theta.
	\end{equation*}
	Here, $ \delta\KI^\mathrm{num}$ is the mode I SIF variation predicted using \cite{lazarus_numerics_2003}'s method. Values of $\epsilon$ for which the errors made by the first- and second-order models exceeds a threshold value are listed in Table~\ref{tab:validation_errors}. Note that we report errors made by our second-order prediction against a numerical estimate of $\KI$, since its exact value is not known. Nonetheless, for the discretization adopted here ($N=\left\lbrace 252, 378, 680, 850\right\rbrace$ for $k=\left\lbrace 2, 3, 4, 5\right\rbrace$, and $\delta a = 0.0025$), the numerical error made in evaluating $\delta \KI^\mathrm{num}$ can be estimated to be less than $0.2\%$. This estimation builds on the error analysis made by \cite{lazarus_numerics_2003} on $\KI$ evaluation along elliptical crack fronts with $b/c=3$ (i.e. $\epsilon=0.66$), that necessitates more small perturbations iterations than here where $\epsilon \leq 0.25$.
	
	\begin{table}
		\centering
		\label{tab:validation_errors}
		\scalemath{0.95}{
			\begin{tabular}{ c c c c c c c }
				& \multicolumn{3}{| c |}{$\vert\vert \cdot \vert\vert_0$} & \multicolumn{3}{| c |}{$\vert\vert \cdot \vert\vert_2$} \\
				& \multicolumn{1}{|c}{1\%} & 5\% & 20\% & \multicolumn{1}{|c}{1\%} & 5\% & \multicolumn{1}{c|}{20\%} \\
				$k=2$ & \multicolumn{1}{|c}{0.2275} & 0.3375 & 0.48 & \multicolumn{1}{|c}{0.3725} & $>$0.5 & \multicolumn{1}{c|}{$>$0.5} \\
				$k=3$ & \multicolumn{1}{|c}{0.1425} & 0.2225 & 0.34 & \multicolumn{1}{|c}{0.2425} & 0.3825 & \multicolumn{1}{c|}{$>$0.5} \\
				$k=4$ & \multicolumn{1}{|c}{0.105} & 0.1675 & 0.265 & \multicolumn{1}{|c}{0.18} & 0.295 & \multicolumn{1}{c|}{0.4875} \\
				$k=5$ & \multicolumn{1}{|c}{0.085} & 0.135 & 0.215 & \multicolumn{1}{|c}{0.1425} & 0.2375 & \multicolumn{1}{c|}{0.405} \\
			\end{tabular}
		}
		\caption{Values of perturbation amplitude $\epsilon$ ($\pm 0.0025$) above which the errors in the SIF variation $\delta \KI$ made by the second-order theory exceeds a certain threshold value $\in \left\lbrace 1\%, 5\%, 25\% \right\rbrace$. Values of $\epsilon$ above $0.5$ have not been investigated in this study.}
	\end{table}
	
	We observe nonetheless that our second-order formula generally yields more accurate predictions than the linear theory of \cite{gao_somewhat_1987}, and that quantitative agreement between the numerical results and the second-order expansion of Eq.~\eqref{eq:cosinus_SIF_variations} is found for $\epsilon \lesssim 0.1$ and $k \in \left[0, 5\right]$. This is further confirmed by the evolution of the average SIF $\left<\KI\right>$ with $\epsilon$ (see Fig.~\ref{fig:validation_cosinus}e):
	\begin{equation}
		\label{eq:cosinus_SIF_average}
		\left<\KI\right> = \dfrac{1}{2\pi} \int_{0}^{2\pi} \KI(\theta) d\theta = \KI^0(a_0) \times \left[1 - \dfrac{k^2 - 4k + 1}{16} \epsilon^2\right] + \mathcal{O}(\epsilon^3), 
	\end{equation}
	and the SIF amplitude $\KI^\mathrm{max}-\KI^\mathrm{min}$ (see Fig.~\ref{fig:validation_cosinus}f):
	\begin{equation}
		\label{eq:cosinus_SIF_amplitude}
		\KI^\mathrm{max}-\KI^\mathrm{min} = \max_{\theta \in \left[0, 2\pi\right]} \KI(\theta) - \min_{\theta \in \left[0, 2\pi\right]} \KI(\theta) = \KI^0(a_0) \times (k-1) \epsilon + \mathcal{O}(\epsilon^3).
	\end{equation}
	The change in average SIF $\left<\KI\right>$ appears indeed of second order in $\epsilon$, and is thus overlooked by the first-order theory. Moreover, we observe a change in the sign of the second-order terms in $\left<\KI\right>$ between $k=3$ and $4$. On the contrary, there is no second-order contribution to the SIF amplitude $\KI^\mathrm{max}-\KI^\mathrm{min}$, and higher-order theories would be required to predict the strong decrease of $\KI$ near $\theta = \pi/k$ observed in Fig.~\ref{fig:validation_cosinus}b. We observed that the abrupt change of slope in the SIF amplitude computed using \cite{lazarus_numerics_2003}'s method is associated with a change in the point where $\KI$ is minimal, from $\theta = \pi/4k$ (see the third panel of  Fig.~\ref{fig:validation_cosinus}b) to $\theta = \pi/k$.
	
	We conclude that (i) our second-order theory generally yields better estimates of the SIF variations along perturbed crack fronts than the first-order model of \cite{gao_somewhat_1987}, but (ii) its accuracy decreases with increasing perturbation amplitude $\epsilon$ and mode $k$. As we will see next, the latter may be rather inconsequential, as the amplitude of the front perturbations is usually a fast decreasing function ($\propto 1/k$) of their mode $k$.
	
	\section{Coplanar crack propagation in weakly disordered materials}
	
	The influence of microscopic fluctuations of toughness on the critical load at which a material may fail has been studied quite extensively for the semi-infinite crack geometry since the seminal study of \cite{roux_effective_2003}. Various methods have been used in the literature to address this problem, like e.g. phase-field models \citep{hossain_effective_2014, brach_anisotropy_2019} or minimal surfaces \citep{ernesti_computing_2021, michel_merits_2022}. But perturbative approaches have played an important role in homogenizing fracture properties. Efforts have been made in both proving the existence of a finite effective toughness for random media \citep{dirr_pinning_2011, dondl_threshold_2020}, and predicting its value from the statistical \citep{roux_self-consistent_2008, patinet_quantitative_2013, demery_microstructural_2014} and spatial \citep{demery_geometry_2014, lebihain_towards_2021} distribution of toughness $\KIc(z,x)$. In particular, it has been shown that the material is toughened by the presence of disorder, i.e. the effective toughness $\KIc^\mathrm{eff}$ is always larger than the average value $\KIc^0$ of the local toughness field. A major pitfall of these studies is that they predict a toughening that scales as the \emph{second-order} of the disorder intensity (measured by e.g. the standard deviation $\sigma$ of the toughness field) from a \emph{first-order} expansion of the stress-intensity factor. Hence, they neglect potential contributions arising from the non-linearities in $\KI$ and may be inaccurate in predicting $\KIc^\mathrm{eff}$.  Moreover, the semi-infinite crack geometry considered in these works also prevents investigating size-effects emerging from the finite length of the crack.
	
	Here, we investigate crack propagation when $\KIc$ is fluctuating randomly within the failure plane $y=0$, and we focus on the effective toughness $\KIc^\mathrm{eff}$ of such composite materials. We propose a rigorous homogenization framework for weakly disordered materials that correctly predicts the non-linear toughening arising from local fluctuations of toughness from a second-order expansion of the stress intensity factor. We also examine the influence of the finiteness of the crack geometry on material toughening, as previous studies mainly focused on the semi-infinite crack geometry. Our analytical predictions are finally compared to the results of $\sim 1$ millions of efficient numerical simulations of crack propagation in heterogeneous brittle media. 
	
	\subsection{Problem statement and field generation}
	
\begin{figure}
		\centering
		\noindent\includegraphics[width=0.45\textwidth]{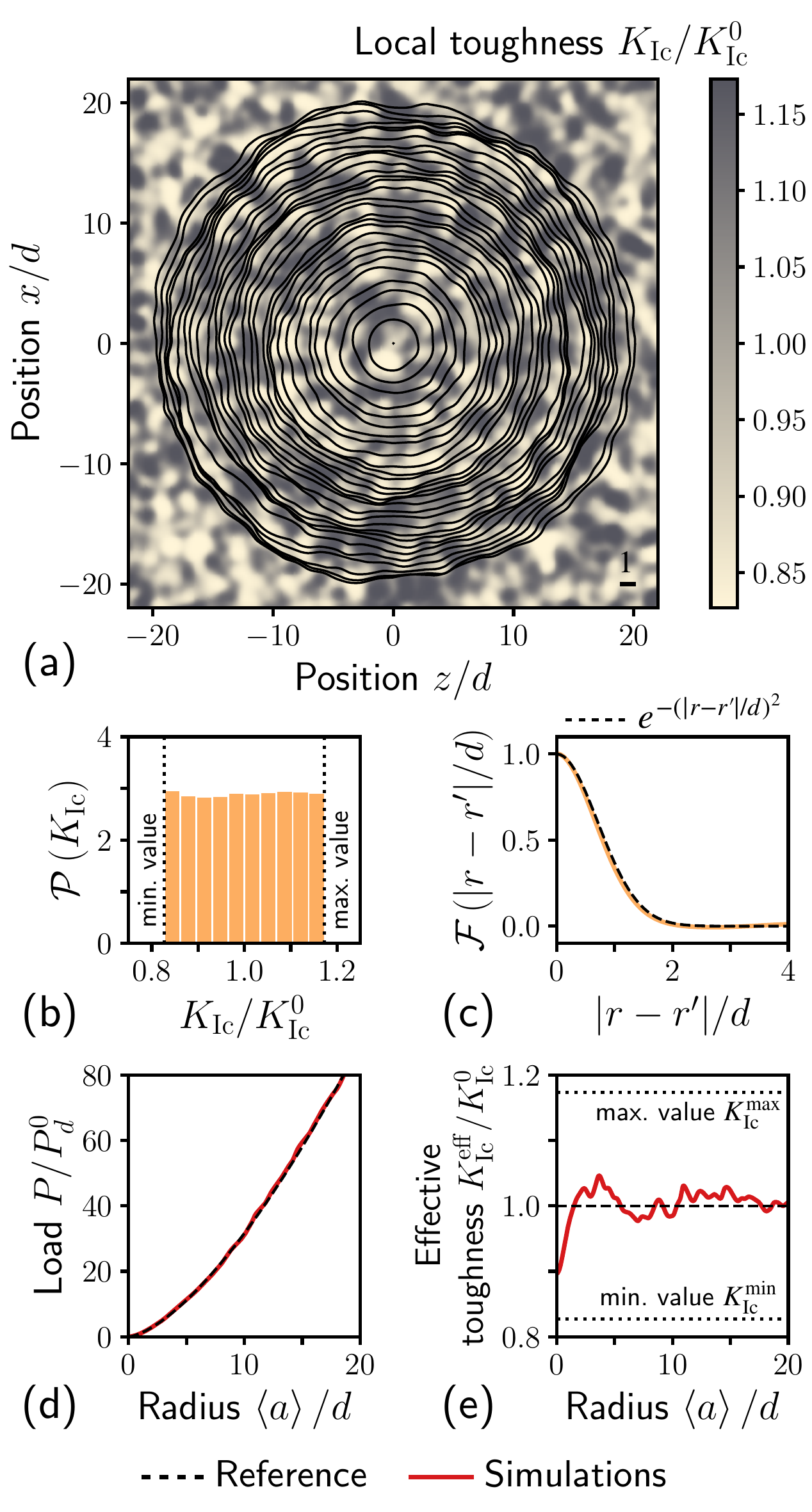}
		\caption{(a) Front positions (in solid lines) of a crack propagating in a disordered toughness field $\KIc$ under the action of a point force $P$ applied at the crack center. The fronts are plotted for a constant load increment. (b) The values of $\KIc$ are uniformly distributed around a mean value $\KIc^0$ with a probability density function $\mathcal{P}(\KIc) = 1/(\KIc^\mathrm{max}-\KIc^\mathrm{min})$, where $\KIc^\mathrm{max}$ and $\KIc^\mathrm{min}$ are the maximum and minimum value of the distribution. (c) The field $\KIc$ is spatially correlated at a typical heterogeneity scale $d$ correlations $\mathcal{F}(\Delta r/d)$ (in solid orange line) following $e^{-(\Delta r/d)^2}$ (in dashed black line). (d) The load $P$ at which the crack propagates (in solid red line), normalized by $P^0_d = \KIc^0 (\pi d)^{3/2}$, fluctuates around its reference value obtained when the toughness field is uniform to $\KIc^0$. (e) As a result, one may estimate an effective toughness $\KIc^\mathrm{eff}$ from the critical loading required to propagate a crack of average radius $\left<a\right>$.}
		\label{fig:disordered_example}
	\end{figure}

	We consider a penny-shape crack of radius $a_\mathrm{ini}$ centered in $O$. The crack propagates along the failure plane located at $y=0$, and interacts with the non-uniform toughness field:
	\begin{equation}
		\label{eq:disordered_toughness}
		\KIc(r, \theta) = \KIc^0 + \sigma f(r, \theta),
	\end{equation}
	where $\KIc^0$ corresponds to the spatial average of the toughness field, $\sigma$ is its standard deviation, and $f$ is a fluctuation term of zero average and unit variance. Fluctuations are either described by the function $f$ in polar coordinates or by $f^*$ in Cartesian coordinates. The two functions are linked through the equation:
	\begin{equation}
		\label{eq:disordered_cartesian_fluctuations}
		f(r, \theta) = f^*(z, x)\text{, where } z=r\cos\theta \text{ and } x=r\sin\theta.
	\end{equation}
	An example of toughness field is given in Fig.~\ref{fig:disordered_example}a. It is characterized its probability density function $\mathcal{P}$ that measures the statistical distribution of the local toughness. Here, we consider that the values of $\KIc$ are uniformly distributed between two extremal values $\KIc^\mathrm{min}$ and $\KIc^\mathrm{max}$ (see Fig.~\ref{fig:disordered_example}b), so that:
	\begin{equation}
		\label{eq:disordered_pdf}
		\mathcal{P}(\KIc) = \dfrac{1}{\KIc^\mathrm{max}-\KIc^\mathrm{min}}.
	\end{equation}
	Consequently, its mean value $\KIc^0$ reads:
	\begin{equation}
		\label{eq:disordered_average}
		\KIc^0 = \dfrac{\KIc^\mathrm{max}+\KIc^\mathrm{min}}{2},
	\end{equation}
	and the intensity $\bar{\sigma} = \sigma/\KIc^0$ of the material disorder is equal to:
	\begin{equation}
		\label{eq:disordered_disorder_intensity}
		\bar{\sigma} = \dfrac{1}{\sqrt{3}} \dfrac{\KIc^\mathrm{max}-\KIc^\mathrm{min}}{\KIc^\mathrm{max}+\KIc^\mathrm{min}}.
	\end{equation}
	The toughness field is further characterized by its spatial structure, described e.g. by the two-points correlation function of $f$. In the following, $\mathrm{E}\left[\cdot\right]$ denotes the expectation of a random variable over multiple disorder realizations. We consider (i) \emph{statistically homogeneous} and (ii) \emph{ergodic} materials, so that ensemble averages (i) do not depend on the observation point $M$, and (ii) are equivalent to spatial average over large enough surfaces. Using these notations, the two-points correlation function of the fluctuations reads:
	\begin{equation}
		\label{eq:disordered_field_correlations}
		\average{f^*(z, x)f^*(z', x')} = \mathcal{F}_z(\abs{z-z'}) \mathcal{F}_x(\abs{x-x'}),
	\end{equation}
	where $\mathcal{F}_z$ and $\mathcal{F}_x$ are the correlation functions in the $z$ and $x$ direction, respectively. Here we assume that the material is isotropic and that the correlations follows a bell-curve, so that:
	\begin{equation}
		\label{eq:disordered_correlations_shape}
		\mathcal{F}_z(u) = \mathcal{F}_x(u) = \mathcal{F}(u) = e^{-u^2/d^2},
	\end{equation}
	where $d$ is the characteristic size of the disorder (see Fig.~\ref{fig:disordered_example}c).
	
	Realizations of $\KIc$ fields are produced following the procedure of \cite{albertini_stochastic_2021}, suitably extended to two-dimensional surfaces. First, a Gaussian random field of zero average, unit variance, and controlled power-spectrum is constructed using the Python package \texttt{FyeldGenerator} \citep{cadiou_fyeldgenerator_2022}. We then apply a non-linear mapping from the Gaussian cumulative distribution function to that of the uniform distribution. This transformation is expected not to affect much the correlation shape $\mathcal{F}$ and the correlation length $d$, as $\mathcal{F}$ is positive \citep{grigoriu_stochastic_2002} (see Fig.~\ref{fig:disordered_example}c).
	
	\subsection{Computation of the crack front position}
	
	The crack is loaded by a pair of symmetric tensile forces of magnitude $P$ applied at the crack center $O$. It gives rise to a stress intensity $\KI^0$ along the front of a penny-shaped crack of radius $a_0$, which reads \citep{tada_stress_2000}:
	\begin{equation}
		\label{eq:pointload_KI_reference}
		\KI^0(a_0) = \dfrac{P}{(\pi a_0)^{3/2}}.
	\end{equation}
	The magnitude $P$ of the tensile forces slowly increases in time, so that the crack extends quasi-statically due to the stabilizing nature of the loading. The successive crack front positions are computed from Irwin's criterion:
	\begin{equation}
		\label{eq:disordered_Irwin}
		\KI(\theta) = \KIc(a(\theta), \theta).
	\end{equation}
	$\KI(\theta)$ is usually computed from the front deformation $\delta a$ from its reference circular configuration of radius $a_0$ using the first-order expansion of \cite{rice_first-order_1985}, while $\KIc(a(\theta), \theta)$ is computed exactly i.e. at all orders in the front deformation. The non-linear Eq.~\eqref{eq:disordered_Irwin} is then solved by building on a viscous regularization of the evolution equation \citep{patinet_quantitative_2013, lebihain_towards_2021}, or by resorting to a comparison theorem to compute efficiently the successive state equilibria \citep{rosso_roughness_2002, demery_geometry_2014}. Here, we follow a different approach, and expand the front position to second-order in the disorder intensity $\bar{\sigma} = \sigma/\KIc^0$:
	\begin{equation}
		\label{eq:disordered_shape}
		a(\theta) = a_0\left[1 + \bar{\sigma} A_1(a_0, \theta) + \bar{\sigma}^2 A_2(a_0, \theta)\right] + \mathcal{O}( \bar{\sigma}^3),
	\end{equation}
	where $a_0$ is the radius of a crack propagating on a homogeneous medium of toughness $\KIc^0$ under the action of the loading $P$. $A_1$ and $A_2$ are fluctuation terms that emerge from the local variations $f$ of the toughness field. One can then expand the local SIF $\KI(\theta)$ at second-order using our Eq.~\eqref{eq:SIF_2ndOrder_general}:
	\begin{equation}
		\label{eq:disordered_SIF_expansion}
		\KI(\theta) = \KI^0(a_0) + \bar{\sigma} \KI^1(a_0, [\KI^0], [a_0 A_1]; \theta) + \bar{\sigma}^2 \left(\KI^1(a_0, [\KI^0], [a_0 A_2]; \theta) + \KI^2(a_0, [\KI^0], [a_0 A_1]; \theta) \right) + \mathcal{O}( \bar{\sigma}^3),
	\end{equation}
	where $\KI^1$ and $\KI^2$ are given by Eq.~\eqref{eq:fourier_SIF_2ndOrder}, with the following convolution kernels:
	\begin{equation}
		\label{eq:pointload_fourier_kernels}
		\begin{cases}
			\begin{aligned}
				& L(k) = - \dfrac{\abs{k}}{2} - \dfrac{3}{2} \\ \vspace{2mm}
				& H(k,k') = \dfrac{R(k,k')}{16} + \dfrac{\abs{k}+\abs{k'}}{4} + \dfrac {3 \abs{k + k'}}{8} + \dfrac{15}{8}.
			\end{aligned}
		\end{cases}
	\end{equation}
	Similarly, the local toughness $\KIc(a(\theta), \theta)$ can be expanded at second-order following:
	\begin{equation}
		\label{eq:disordered_toughness expansion}
		\KIc(a(\theta), \theta) = \KIc^0 \left( 1 + \bar{\sigma} f(a_0, \theta) + \bar{\sigma}^2 \dfrac{\partial f}{\partial r}(a_0, \theta) a_0 A_1(a_0, \theta) \right).
	\end{equation}
	The crack front shape is then computed by solving Irwin's criterion of Eq.~\eqref{eq:disordered_Irwin} in cascade, i.e. at successive but increasing orders in $\bar{\sigma}$. Starting with zero order, one gets from Eq.~\eqref{eq:pointload_KI_reference}:
	\begin{equation}
		\label{eq:disordered_Irwin_0thOrder}
		\KI^0(a_0) = \KIc^0 \Rightarrow a_0 = \dfrac{1}{\pi} \left(\dfrac{P}{\KIc^0}\right)^{2/3},
	\end{equation}
	which sets the instantaneous value of the loading $P$. One could have alternatively set the magnitude $P$ of the opening forces, and solve Eq.~\eqref{eq:disordered_Irwin_0thOrder} to find the reference radius $a_0$. At first order, one has:
	\begin{equation}
		\label{eq:disordered_Irwin_1stOrder}
		\KI^1(a_0, [\KI^0], [a_0 A_1]; \theta) = \KIc^0 f(a_0, \theta) \Rightarrow \sum_k L(k) \hat{A}^1_k(a_0) e^{ik\theta} = \sum_k \hat{f}_k(a_0) e^{ik\theta},
	\end{equation}
	so that the $k$-th Fourier coefficient of $A_1$ reads:
	\begin{equation}
		\label{eq:disordered_A1_fourier}
		\hat{A}^1_k(a_0) = \dfrac{\hat{f}_k(a_0)}{L(k)},
	\end{equation}
	where $L(k)$ is given in Eq.~\eqref{eq:pointload_fourier_kernels}. We observe here that the perturbation amplitude is a fast decreasing function of its mode $k$ ($\propto 1/k$), and that crack fronts are generally stiffer to small wavelength perturbations. Collecting the second-order terms in Eq.~\eqref{eq:disordered_Irwin}, one finds:
	\begin{equation}
		\label{eq:disordered_Irwin_2ndOrder}
		\begin{aligned}
			& \KI^1(a_0, [\KI^0], [a_0 A_2]; \theta) + \KI^2(a_0, [\KI^0], [a_0 A_1]; \theta) = \KIc^0 a_0 \dfrac{\partial f}{\partial r}(a_0, \theta) A_1(\theta)\\
			\Rightarrow\, & \sum_k L(k) \hat{A}^2_k(a_0) e^{ik\theta} = \sum_{k, k'} a_0\widehat{\left(\dfrac{\partial f}{\partial r}\right)}_{k'}\,(a_0) \hat{A}^1_{k-k'}(a_0) e^{ik\theta} - \sum_{k, k'} H(k', k-k') \hat{A}^1_{k'}(a_0) \hat{A}^1_{k-k'}(a_0) e^{ik\theta},
		\end{aligned}
	\end{equation}
	where $H(k,k')$ is given in Eq.~\eqref{eq:pointload_fourier_kernels}. We observe that each mode $\hat{A}^2_k$ of the second-order fluctuations $A_2$ are linked to all perturbation modes of $A_1$. Injecting Eq.~\eqref{eq:disordered_A1_fourier}, the $k$-th Fourier coefficient of $A_2$ reads:
	\begin{equation}
		\label{eq:disordered_A2_fourier}
		\hat{A}^2_k(a_0) = \dfrac{1}{L(k)} \sum_{k'} \left(\dfrac{a_0}{L(k-k')}\widehat{\left(\dfrac{\partial f}{\partial r}\right)}_{k'}\,(a_0) \hat{f}_{k-k'}(a_0) - \dfrac{H(k', k-k')}{L(k')L(k-k')}\hat{f}_{k'}(a_0)\hat{f}_{k-k'}(a_0)\right).
	\end{equation}
	In practice, the crack front shape $a(\theta)$ and the critical loading $P$ are computed using Eqs.~\eqref{eq:disordered_shape}, \eqref{eq:disordered_Irwin_0thOrder}, \eqref{eq:disordered_A1_fourier}, and \eqref{eq:disordered_A2_fourier}. For a given reference crack radius $a_0$, the magnitude of the applied force $P$ is computed from Eq.~\eqref{eq:disordered_Irwin_0thOrder}. The crack front is then discretized in $n=1024$ intervals of equal length, and Eqs.~\eqref{eq:disordered_A1_fourier}, and \eqref{eq:disordered_A2_fourier} are solved using fast Fourier Transforms. The computation time for one reference front position $a_0$ scales as $\mathcal{O}(n^2\ln n)$. Examples of crack front shapes and loading evolution are given in Fig.~\ref{fig:disordered_example}a and \ref{fig:disordered_example}d.
	
	Our numerical approach differs from those available in the literature, as only second-order contributions of the toughness fluctuations are taken into account in coherence with the second-order expansion of $\KI$. We expect it to break as soon as the front deformation $\delta a(\theta) = a(\theta)-a_0$ gets larger than the characteristic length scale $d$ of the toughness variations. Indeed, for larger front distortions, the second-order expansion of the local toughness in Eq.~\eqref{eq:disordered_toughness expansion} may not be valid anymore. This strongly limits (i) the intensity $\bar{\sigma}$ of the material disorder, and (ii) the ratio $a_0/d$ between the crack radius to heterogeneity size considered in the simulations. Namely, we measured from our numerical computations that we could not model crack radii larger than $a_0 = 20d$ for $\bar{\sigma}=0.1$.
	
	\subsection{Effective toughness of weakly disordered materials}
	
	\review{Due to the strong decrease of $\KI^0$ with crack advance (see Eq.~\eqref{eq:pointload_KI_reference}), we do not observe any snapback instability in the evolution of $P$ with the average crack radius $\left<a\right>$ (see Fig.~\ref{fig:disordered_example}d), which reads:}
	\begin{equation}
		\label{eq:disordered_average_position}
		\left<a\right>/a_0 = 1+ \bar{\sigma} \hat{A}^1_0(a_0) + \bar{\sigma}^2 \hat{A}^2_0(a_0) + \mathcal{O}( \bar{\sigma}^3).
	\end{equation}
	As a result, all states computed from Eqs.~\eqref{eq:disordered_Irwin_0thOrder}, \eqref{eq:disordered_A1_fourier}, and \eqref{eq:disordered_A2_fourier} are \review{state equilibria}. This suggests that the instability nuclei, often referred to as Larkin domains, are always larger than the crack perimeter $\mathcal{P} \simeq 2\pi \left<a\right>$. \review{Accordingly, our simulations should fall down in the \emph{weak pinning} regime identified by \cite{roux_effective_2003} for small disorder intensity $\bar{\sigma}$. In this regime, the material toughening:}
	\begin{equation}
		\label{eq:disordered_toughening}
		T = \KIc^\mathrm{eff}/\KIc^0 - 1
	\end{equation}
	\review{is expected to be zero \citep{roux_effective_2003}. Our goals here are (i) to show that this is wrong if one considers a second-order expansion of both the stress-intensity factor and the toughness field, and (ii) to assess the dependence of the observed toughening on the crack size $a_0$.}\\
	
	At the macro-scale, one may overlook the undulations of the crack front and consider the perturbed crack as a circular crack of radius $\left<a\right>$ (see Fig.~\ref{fig:disordered_example}a), which departs from its reference value $a_0$ under the action of the material disorder (see Eq.~\eqref{eq:disordered_average_position}). As the crack propagates under the action of the tensile forces, one may then define an apparent, or \emph{effective}, toughness of the heterogeneous media from the magnitude $P$ of the tensile forces required to make a circular crack of radius $\left<a\right>$ propagate:
	\begin{equation}
		\label{eq:disordered_effective_toughness}
		\KIc^\mathrm{eff} = \dfrac{P}{(\pi \left<a\right>)^{3/2}},
	\end{equation}
	so that:
	\begin{equation}
		\KIc^\mathrm{eff} =\KI^0(a_0) \left(\dfrac{a_0}{\left<a\right>}\right)^{3/2} =\KIc^0 \left(\dfrac{a_0}{\left<a\right>}\right)^{3/2}.
	\end{equation}
	We observe that the toughening at the macroscale emerges from the difference between the average front position $\left<a\right>$ and its reference value $a_0$. If the front is pinned by the disorder ($\left<a\right> < a_0$), the composite appears tougher at the macroscale than the reference homogeneous material of toughness $\KIc^0$. On the contrary, if the disorder facilitates crack propagation ($\left<a\right> > a_0$), the material is weakened by the heterogeneities. Using Eq.~\eqref{eq:disordered_average_position}, one finds:	
	\begin{equation}
		\label{eq:disordered_normalized_toughness}
		\KIc^\mathrm{eff}/\KIc^0 = \KI^0(\left<a\right>)/\KI^0(a_0) = 1 + \bar{\sigma} L(0) \hat{A}^1_0(a_0) + \bar{\sigma}^2\left(L(0)\hat{A}^2_0(a_0) + H(0,0)\hat{A}^1_0(a_0)^2\right),
	\end{equation}
	where \review{$L(0) = a_0 (\partial \KI^0/\partial a)(a_0)/\KI^0(a_0)$ and $H(0,0) = a_0^2 (\partial^2 \KI^0/\partial a^2)(a_0)/2\KI^0(a_0)$}. An example of the evolution of $\KIc^\mathrm{eff}$ with the average crack radius $\left<a\right>$ is given in \review{Fig.~\ref{fig:disordered_example}e}. 
	
	We observe in Fig.~\ref{fig:disordered_example}d that the effective toughness is a random variable that seems to fluctuate around $\KIc^0$ and between the two extremal values $\KIc^\mathrm{min}$ and $\KIc^\mathrm{max}$ of the toughness distribution. To have a better perspective on the statistical distribution of the random variable $\KIc^\mathrm{eff}$, we perform more than $1$ million Monte-Carlo simulations of crack propagation in heterogeneous media, following the procedure described in Section~4.2. Namely, the disorder intensity is varied between 11 different values $\bar{\sigma} \in \left[0.001, 0.01\right]$. $N=100'000$ independent and identically distributed realizations of the toughness field $\KIc(z,x)$ are generated for each value of $\bar{\sigma}$. For each disorder realization, the crack front position is computed from 16 reference radius $a_0 \in \left[0.02\,d,20\,d\right]$. It results on $17.6$ millions of crack fronts from which one can estimate the average front position $\left<a\right>$ using Eq.~\eqref{eq:disordered_average_position}, and the effective toughness $\KIc^\mathrm{eff}$ following Eq.~\eqref{eq:disordered_normalized_toughness}. 
	
	\begin{figure}
		\begin{minipage}{0.45\textwidth}
			\centering
			\includegraphics[width=\textwidth]{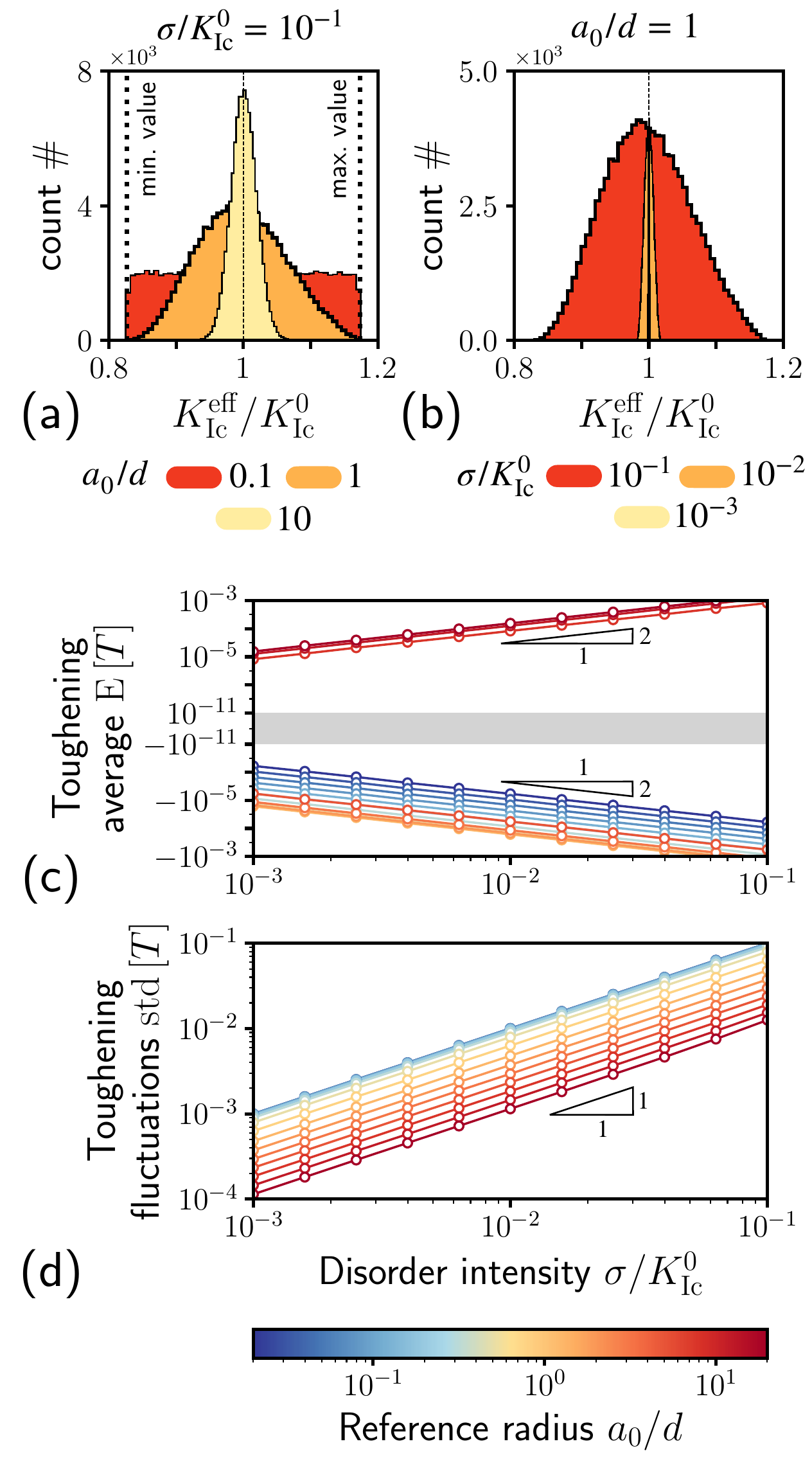}
			\caption{(a) At constant disorder intensity $\sigma/\KIc^0 = 10^{-1}$, the simulated values of effective toughness $\KIc^\mathrm{eff}$ are distributed between the two extremal values $\KIc^\mathrm{min}$ and $\KIc^\mathrm{max}$ of the local toughness distribution, and its distribution depends on the ratio between the reference crack radius $a_0$ and the heterogeneity size $d$. (b) Similarly, at constant size ratio $a_0/d = 0.1$, the distribution of $\KIc^\mathrm{eff}$ depends on the disorder intensity $\sigma/\KIc^0$ (the distribution plotted in bold is the same to that shown in \ref{fig:disordered_statistics}a). As a result, (c) the average $\mathrm{E}\left[T\right]$ and (d) the standard deviation $\mathrm{std}\left[T\right]$ of the toughening $T = \KIc^\mathrm{eff}/\KIc^0-1$ depends on both $\sigma/\KIc^0$ and $a_0/d$. In particular, $\mathrm{E}\left[T\right]$ scales as $(\sigma/\KIc^0)^2$ and its sign depends on $a_0/d$, while $\mathrm{std}\left(T\right)$ is linear in $\sigma/\KIc^0$.}
			\label{fig:disordered_statistics}
		\end{minipage}%
		\hspace{0.1\textwidth}
		\begin{minipage}{0.45\textwidth}
			\centering
			\includegraphics[width=\textwidth]{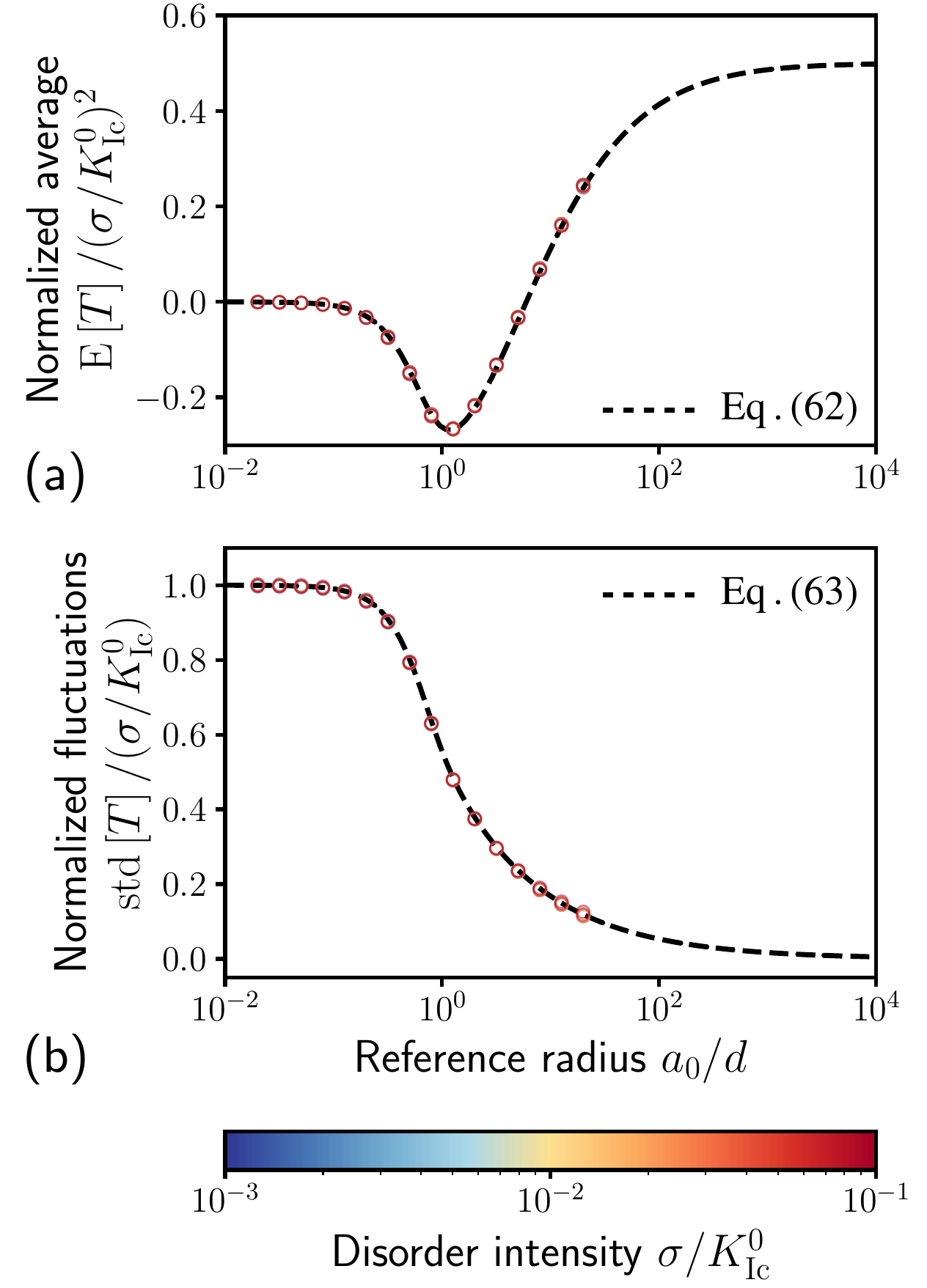}
			\caption{(a) When normalized by the square of the disorder intensity $(\sigma/\KIc^0)^2$, the values of the average $\mathrm{E}\left[T\right]$ of the toughening showed in Figure~\ref{fig:disordered_statistics}c all fall down onto a master curve predicted by Eq.~\eqref{eq:disordered_toughening_average}. (b) Similarly, when normalized by $(\sigma/\KIc^0)$, the values of the standard deviation $\mathrm{std}\left[T\right]$ presented in Figure~\ref{fig:disordered_statistics}d follow the master curve of Eq.~\eqref{eq:disordered_toughening_std} controlled by $a_0/d$.}
			\label{fig:disordered_scalings}
			\vspace{4cm}
		\end{minipage}
	\end{figure}
	
	The results of our statistical analysis are summarized in Fig.~\ref{fig:disordered_statistics}. For $\bar{\sigma}=0.1$, we observe that the effective toughness is indeed bounded by the two extremal values $\KIc^\mathrm{min}$ and $\KIc^\mathrm{max}$ of the toughness distribution (see Fig.~\ref{fig:disordered_statistics}a), as expected from Irwin's criterion. For a small crack radius $a_0 = 0.1\,d$, the effective toughness follows the probability distribution of the local toughness $\KIc$ (see Fig.~\ref{fig:disordered_example}b), as the crack front is often embedded into one single heterogeneity. As the crack grows larger, the distribution of the effective toughness departs from that of the local toughness and strongly resemble a Gaussian distribution. Interestingly, the variance of the distribution looks to decrease with increasing crack to heterogeneity size ratio $a_0/d$. One can also notice that the average value of the toughness distribution looks slightly below $\KIc^0$ for $a_0/d=1$, and closer to it for $a_0/d=10$. Departure of this average value from $\KIc^0$ seems in general very small with respect to the fluctuations of $\KIc^\mathrm{eff}$, so that both quantities may not be of the same order in $\bar{\sigma}$. We also observe in Fig.~\ref{fig:disordered_statistics}b that the fluctuations of the effective toughness increase with the disorder intensity $\bar{\sigma}$.\\
	
	In the following, we do not try to predict the exact probability distribution of $\KIc^\mathrm{eff}$ in Fig.~\ref{fig:disordered_statistics}a-b, but rather focus on its statistical descriptors, namely its average value and its standard deviation. As the values of $\KIc^\mathrm{eff}$ seems to gather around $\KIc^0$, we concentrate instead on the statistics of the toughening $T=\KIc^\mathrm{eff}/\KIc^0-1$. By performing ensemble averages on Eq.~\eqref{eq:disordered_normalized_toughness}, one can show that its average reads:
	\begin{equation}
		\label{eq:disordered_toughening_average}
		\average{T} = \left(\dfrac{\sigma}{\KIc^0}\right)^2 \left[\sum_{k} \left(\dfrac{1}{L(k)} \hat{G}_{k}(a_0/d) - \dfrac{H(k,-k)}{L(k)^2} \hat{F}_{k}(a_0/d)\right) + \dfrac{H(0,0)}{L(0)^2}\hat{F}_0(a_0/d) \right] + \mathcal{O}\left(\left(\dfrac{\sigma}{\KIc^0}\right)^3\right),
	\end{equation}
	while its standard deviation writes as:
	\begin{equation}
		\label{eq:disordered_toughening_std}
		\std{T} = \left(\dfrac{\sigma}{\KIc^0}\right) \hat{F}_0(a_0/d)^{1/2} + \mathcal{O}\left(\left(\dfrac{\sigma}{\KIc^0}\right)^2\right),
	\end{equation}
	where $F$ and $G$ are the functions:
	\begin{equation}
		\label{eq:disordered_scaling}
		F(a_0/d, \theta) = e^{-4(a_0/d)^2\sin^2(\theta/2)} \text{, and } G(a_0/d, \theta) = -4(a_0/d)^2 \sin^2(\theta/2) e^{-4(a_0/d)^2\sin^2(\theta/2)},
	\end{equation}
	which both relate to the correlation shape $\mathcal{F}$ of the fluctuations $f$. Details on the derivation of Eqs.~\eqref{eq:disordered_toughening_average} and \eqref{eq:disordered_toughening_std} are given in \ref{apdx:statistics}. Several comments are in order:
	\begin{itemize}
		\item We observe that the average toughening $\average{T}$ scales as $\bar{\sigma}^2$, while $\std{T}$ increases linearly with $\sigma$. As a consequence, it is extremely difficult to measure $\average{T}$ from our ``naive'' Monte-Carlo simulations, as the central limit theorem states that the fluctuations of $\average{T}$ are proportional to $\bar{\sigma}/\sqrt{N}$, where $N$ is the number of realizations. To circumvent this issue, we use a variance reduction technique, and resort to antithetic variates. Let us call $\Omega_1$ the ensemble containing all $N=100'000$ realizations generated here for fixed values of $\bar{\sigma}$ and $a_0/d$. We attribute to each realization an integer $i$, and note $T_1^i$ the value of $T$ measured at $a_0$ for a crack propagating in the toughness field $\KIc^i = \KIc^0 (1+\bar{\sigma}f_i)$. We now consider the ensemble $\Omega_2$ containing the 100'000 realizations corresponding to the toughness field $\bar{\KIc}^i= \KIc^0 (1-\bar{\sigma}f_i)$ (opposite fluctuations), and note $T_2^i$ the value of $T$ measured at $a_0$ in this case. $T_2^i$ can be computed at almost no cost, as it corresponds to a change of sign in front of $A_1$ only. As both $f_i$ and $-f_i$ are identically distributed (zero average, unit variance, uniform probability distribution, same two-points correlation function), one has:
		\begin{equation}
			\average{T} = \dfrac{\average{T_1}+\average{T_2}}{2} \simeq \dfrac{ \sum_{i \in \Omega_2} T_1^i + \sum_{i \in \Omega_2} T_2^i}{2N},
		\end{equation}
		but one expects that the covariance of the random variables $T_1$ and $T_2$ is negative, so that the variance of $\tilde{T} = (T_1+T_2)/2$ is greatly reduced. Using antithetic variates, one observe from Fig.~\ref{fig:disordered_statistics}c that the average toughening $\average{T}$ do scales as $\bar{\sigma}^2$, while its standard deviation $\std{T}$ is only linear in $\bar{\sigma}$ (see Fig.~\ref{fig:disordered_statistics}d). Note that numerical validations of these scalings can only be obtained through the use of variance reduction methods, as no convergence was found even for $N = 100'000$. Using our antithetic variates, one observes satisfactory scaling for $N$ as low as $N \simeq 250$. Our trick may find some use if one wanted to compare the results of perturbative calculations to that of e.g. BEM simulations.
		
		\item \review{In contrast with the predictions of \cite{roux_effective_2003}, \cite{patinet_quantitative_2013}, and \cite{demery_microstructural_2014}, the effective toughness $\KIc^\mathrm{eff}$ is not equal to the spatial average $\KIc^0$ of the toughness field while being in the weak pinning regime. In the semi-infinite crack limit ($a_0/d \rightarrow +\infty$), the average toughening $\average{T}$ is equal to $\average{T} = 0.5 (\sigma/\KIc^0)^2$ (see Fig.~\ref{fig:disordered_scalings}a). This toughening contribution is dictated by non-linearities in $\KI$, as only the terms arising from the second-order expansion of $\KI$ actually play a role in determining $\average{T}$ (through the terms involving the convolution kernel $H$ in Eq.~\eqref{eq:disordered_toughening_average}). Moreover, the toughening predicted here in the \emph{weak pinning} regime is comparable to that measured by \cite{demery_microstructural_2014} and \cite{demery_geometry_2014} in the \emph{strong pinning} regime, where crack propagation is no more smooth and occurs as the succession of depinning instabilities \citep{roux_effective_2003}. Indeed, the predictions $\average{T}$ of Eq.~\eqref{eq:disordered_toughening_average} for a semi-infinite crack amounts to one fourth of the toughening measured by \cite{demery_microstructural_2014} (see their Eq. (20)), and half that predicted by \cite{demery_geometry_2014} (see their Eqs. (34) and (35)). We conclude that one cannot neglect non-linearities in the stress intensity factor variations when estimating the effective toughness of brittle composites.}
		
		This calls for a new derivation of the theoretical predictions of \cite{patinet_quantitative_2013} and \cite{demery_microstructural_2014} for a semi-infinite crack in the strong pinning regime, involving now the second-order terms in the expansion of $\KI$ or $G$ derived by \cite{vasoya_second_2013}. \review{To do so, one can follow the route proposed by \cite{demery_geometry_2014}, which is based on a second-order expansion of the front position similar to ours, but taking into account a finite driving velocity. These predictions should then be compared to the results of simulations performed in the strong pinning regime, in which the crack jumps from one equilibrium position to the other. One could build on the approach recently proposed by \cite{sanner_crack-front_2022}, and derive a variational formulation of the problem using a third-order expansion of the elastic energy. The successive state equilibria can then be computed from a trust-region Newton conjugate-gradient algorithm.} Another promising route would be to extend the numerical method of \cite{rosso_roughness_2002} to second-order, but one would need first to assess the validity of Middleton's comparison theorem for the non-linear convolution integrals of \cite{vasoya_second_2013}.
		
		\item We observe in Fig.~\ref{fig:disordered_statistics}c and Fig.~\ref{fig:disordered_scalings}a that the presence of material disorder does not always result in an overall toughening of the material, in contrast with the results obtained for the semi-infinite crack \citep{patinet_quantitative_2013, demery_microstructural_2014, lebihain_towards_2021}. Indeed, for very small crack radii $a_0 \ll d$, the average toughening $\average{T}$ is zero, as the crack propagates in a single heterogeneity of uniform toughness, whose probability distribution is given by $\mathcal{P}$ of Eq.~\eqref{eq:disordered_pdf} and average value is $\KIc^0$. For larger but still small crack radii $a_0 \lesssim 6d$, the average toughening $\average{T}$ is negative. The presence of material disorder leads to an overall weakening of the composite from its average toughness $\KIc^0$. On the contrary, the material is reinforced by it for larger crack radii $a_0$, in agreement with the results obtained for the semi-infinite crack. These variations are accurately captured by Eq.~\eqref{eq:disordered_toughening_average}. We show in Fig.~\ref{fig:disordered_scalings}a the normalized average $\average{T}/\bar{\sigma}^2$ of the toughening as measured in our numerical simulations. All data points collapse to the master curve described by Eq.~\eqref{eq:disordered_toughening_average}. We further observe from Fig.~\ref{fig:disordered_scalings}a that the average toughening $\average{T}$ saturates in the limit of very large crack radii $a_0 \gg d$. This increase of the macroscopic toughness with crack length up to a saturation value bears striking similarities with the R-curve behavior observed in a wide variety of materials like e.g. concrete \citep{bazant_r-curve_1993}, ceramics \citep{ebrahimi_slow_2000}, rocks \citep{funatsu_combined_2004}, bone \citep{nalla_mechanistic_2005}, compacted granular materials \citep{girardot_disk-shaped_2023}, etc.
		
		A natural consequence of this R-curve behavior is that material disorder may promote micro-cracking below $\simeq d$. Indeed, the extension of pre-existing microscopic defects is facilitated by the overall decrease of the effective toughness up to $a_0 = d$, while their growth is further stabilized by the apparent increase of the fracture resistance. Heterogeneous materials may thus display an increased ductility at the macroscale.
		
		\item We see in Fig.~\ref{fig:disordered_scalings}b that the fluctuations of the toughening $T$ (or equivalently that of the effective toughness $\KIc^\mathrm{eff}$) decreases with increasing crack radius $a_0$. The dependence of $\std{T}$ in $a_0/d$ is well captured by our theory, as one can see from Fig.~\ref{fig:disordered_scalings}b, where numerical estimates of the normalized fluctuations $\std{T}/\bar{\sigma}$ all collapse to the master curve described by Eq.~\eqref{eq:disordered_toughening_std}. Again, one notices that $\std{\KIc^\mathrm{eff}} = \sigma = \std{\KIc}$ for small crack radii $a_0/d \ll 1$, as the crack propagates in a single heterogeneity only. \review{Moreover, for large crack radii $a_0/d \gg 1$, $\std{T} \propto \sigma (a_0/d)^{-1/2}$. This finite size effect may be reminiscent of having Larkin domains smaller than the front perimeter \citep{demery_microstructural_2014}.} As $\std{\KIc^\mathrm{eff}} \underset{a_0/d \rightarrow +\infty}{\rightarrow} 0$, it is tempting to think that our theory proves the existence of an \emph{intrinsic} (deterministic) effective toughness for weakly disordered brittle composites in the limit $a_0/d \rightarrow +\infty$. However, one must keep in mind that Eq.~\eqref{eq:disordered_toughening_std} only provides a first-order expansion of the standard deviation $\std{T}$ of the toughening, so that the second-order fluctuations of the effective toughness are not necessarily zero in that limit. To access these second-order fluctuations, one should derive a third-order expansion of the variance $\Var{T}$ of the toughening. Note that this is feasible here, even within a second-order expansion of $\KI$, but it would involve the three-points correlation function of $f$, a quantity that is often poorly constrained and difficult to compute \citep{jiao_superior_2009}. Those calculations are out-of-scope of the present study, and are left for further works. 
		
		\item The expressions of the average of the toughening $\average{T}$ and its standard deviation $\std{T}$ involve the functions $F$ and $G$ that relate to the two-points correlation function $\mathcal{F}$ of the toughness fluctuations $f$. This was already noticed by \cite{demery_geometry_2014} for the semi-infinite crack, and it suggests that playing on the spatial structure of the toughness field could potentially lead to a change in material toughening. An interesting route would be to investigate cases where the correlation length $d_z$ of the toughness field in the direction $z$ is different from that $d_x$ in the direction $x$. One could then expand the front shape $a(\theta)$ with the disorder intensity $\bar{\sigma}$, but also with a small parameter $\epsilon = d_x/d_z-1$ that quantifies the toughness anisotropy. One could then observe how it modifies both the average front shape, and the overall toughening of the material. Note that such an expansion is only possible within our second-order theory, as terms of first-order in $\epsilon$ are necessarily zero (absence of disorder).
		
		\item $\average{T}$ involves sums over the Fourier convolution kernels $L(k)$ and $H(k,k')$. As the latter both depend the reference SIF $\KI^0$ generated by the loading, and its derivatives $\partial \KI^0/\partial a$ and $\partial^2 \KI^0/\partial a^2$ with crack advance (see Eq.~\eqref{eq:fourier_kernels_2ndOrder}), one expects that the effective toughness $\KIc^\mathrm{eff}$ generally depends on the loading conditions. This was already noted by \cite{demery_geometry_2014} and \cite{lebihain_effective_2021} in the case of the semi-infinite crack. This dependency breaks down as soon as the structural length scales set by both the crack size $a_0$ and the loading conditions:
		\begin{equation}
			\label{eq:structural_lengthscales}
			\mathcal{L}_1 = \KI^0/(d\KI^0/da) \text{ and } \mathcal{L}_2 = \sqrt{\KI^0/(d^2\KI^0/da^2)}.
		\end{equation}
		get much larger than the heterogeneity size $d$. Note however that the standard deviation of the toughening $\std{T}$ given in Eq.~\eqref{eq:disordered_toughening_std} is independent of the loading conditions.
	\end{itemize}
	
	Overall, our findings shed new lights on the toughening of brittle composites by heterogeneities, as well as on the size effects emerging from the finiteness of the crack geometry. However, one must keep in mind that they are only valid as long as a second-order expansion of $\KIc$ remains valid, i.e. when the front perturbation $\delta a$ is smaller than the characteristic length scale $d$ of the toughness field ($\delta a \lesssim d$). \review{As a result, one should expect departure of the numerical results and our theory as soon as the crack enters the so-called \emph{strong pinning} regime, in which crack propagation articulates as the succession of depinning instabilities \citep{demery_microstructural_2014}. But one should also be aware that taking into account size effects and non-linearities in the strong pinning regime may change the conclusions obtained by \cite{patinet_quantitative_2013} and \cite{demery_microstructural_2014}.}
	
	\section{Conclusion}
	\label{sec:conclusion}
	
	This paper address the second-order variation of the local mode I stress intensity factor (SIF) arising from a coplanar perturbation of the front of a penny-shaped crack embedded in some isotropically linearly elastic infinite body. We build next on our model to determine material reinforcement arising from local fluctuations of toughness, and account for both non-linear and finite-size effects.
	
	Section 2 was devoted to the derivation of our theoretical model, following the approach proposed by \cite{rice_weight_1989} and used by \cite{leblond_second_2012} for the semi-infinite crack. First, we derived the first-order variations of the fundamental kernel $Z$ arising from the perturbation of the crack front from its reference circular configuration. Next, we determined the first-order variations of the derivative $\partial\KI/\partial A$ of the mode I SIF $\KI$ with respect to the amplitude $A$ of the perturbation, which ultimately yields our second-order formula of Eqs.~(\ref{eq:fourier_SIF_2ndOrder}-\ref{eq:fourier_kernels_2ndOrder}) by integration. This was performed in the case where the unperturbed crack is loaded by some axisymmetric system of forces, but the general derivation was proposed in \ref{apdx:calculation_K}.
	
	In Section 3, we validated our second-order theory by comparing its predictions to the second-order expansion of an analytical formula derived by \cite{irwin_fracture_1958} for the SIF distribution along the front of an elliptical crack loaded by a uniform tensile stress. The validity range of our model was then assessed by comparing its output to the local SIF distributions along wavy cracks of varying mode and amplitude, which were computed using the numerical method of \cite{lazarus_numerics_2003}. We showed that the predicting abilities of our second-order theory are improved from the linear model of \cite{gao_somewhat_1987}, but that errors are increasing with both the perturbation mode and amplitude.
	
	In Section 4, we use our non-linear model to investigate crack propagation in heterogeneous media, in which the local toughness was fluctuating at random. \review{We only considered moderate fluctuations of material toughness, so that crack propagation was stable and occurred as the continuous succession of equilibrium positions (weak pinning regime). The toughening arising from toughness fluctuations has been widely investigated in the literature for the semi-infinite coplanar crack \citep{roux_effective_2003}, even beyond the case of weak pinning, \citep{patinet_quantitative_2013, demery_microstructural_2014, lebihain_towards_2021}. However, these authors predicted a zero toughening for weak pinning, building a \emph{first-order} expansion of the local stress intensity factor $\KI$. Here, we showed that this was no more true as soon as one considers non-linearities in the expression of $\KI$.} More precisely, we showed that the effective toughness of disordered brittle media is a random variable, whose statistical descriptors, like e.g. its average and its variance, can be predicted rigorously at second order in $\bar{\sigma}$ using our non-linear theory. We validated our theoretical predictions on the results of $\sim 1$ million numerical simulations of crack propagation in weakly disordered materials. One of our key findings is that material toughening by heterogeneities of typical size $d$ is \emph{size-dependent}. Namely, cracks of average radius $\left<a\right>$ smaller than $\simeq 6\,d$ are experiencing an effective toughness $\KIc^\mathrm{eff}$ that is lower than the average $\KIc^0$ of the toughness field. On the contrary, larger cracks propagates in a seemingly tougher material ($\KIc^\mathrm{eff} > \KIc^0$), in accordance with previous results obtained for the semi-infinite crack. It leads to an apparent R-curve behavior of the brittle composite at the macroscale. A natural consequence of this observation is that material disorder may promote the development of micro-cracks at the heterogeneity scale, while preventing them to extend further, leading to an overall increase in material ductility. We also proved that material toughening scales, on average, as the square $\bar{\sigma}^2$ of the fluctuations amplitude, and that non-linear terms in the expansion of $\KI$ do play a role in the expression of the associated prefactor. \review{Moreover, this contribution in toughening is comparable to that emerging from the intermittent dynamics of crack front at moderate disorder intensity \citep{demery_geometry_2014}. This drive the needs to revisit the results of \cite{patinet_quantitative_2013} and \cite{demery_microstructural_2014} in a non-linear setting, but appropriate numerical methods must  be developed to validate theoretical predictions.}
	
	Our work also raises interesting perspectives to model crack propagation in a multi-physics environment. First, we provided here a second-order expansion of the stress intensity factor along quasi-circular crack fronts. However, the perturbative approach of \cite{rice_weight_1989} also yields the first-order variations of local aperture, a quantity that has rarely been used in the past. One could use a reasoning similar to that of \cite{leblond_second_2012}, and derive its second-order expansion. Combined with that of the crack face weight function, it should provide means to investigate crack propagation in heterogeneous yet cohesive materials, whose strength $\sigma_\mathrm{c}$ depends on the local crack opening \citep{lebihain_cohesive_2022}. The knowledge of the aperture of perturbed cracks also gives access to the crack volume, which may be relevant to describe 3D hydraulic fracturing \citep{savitski_propagation_2002} and its interplay with material heterogeneities. Second, one could extend our mode I calculations to mixed mode I+II+III using the perturbative framework of \cite{favier_coplanar_2006}. Preliminary calculations show that it should allow rationalizing the quasi-elliptical front shapes of fluid-driven shear ruptures propagating along frictional interfaces governed by Coulomb's friction \citep{saez_three-dimensional_2022}. Third, the knowledge of the perturbation $\delta a_*$, which has been derived in Section~2.1 to determine the first-order variation of the fundamental kernel, permits us to extend our second-order model to nearly circular connections \citep{gao_nearly_1987} at relatively small cost. One may then apply this theory to investigate adhesion along chemically heterogeneous interfaces \citep{sanner_crack-front_2022} or contact between a soft material and a rough indenter \citep{argatov_controlling_2021}.
	
	\section*{CRediT authorship contribution statement}
	\noindent
	\textbf{Mathias Lebihain:} Conceptualization, Investigation, Formal analysis, Numerical simulations, Software, Visualization, Writing - Original Draft.
	\textbf{Manish Vasoya:} Investigation, Numerical simulations, Writing - Review \& Editing.
	\textbf{V{\'e}ronique Lazarus:} Investigation, Software, Writing - Review \& Editing.
	
	\section*{Online access to data}
	The Python script used to generate the figures and the associated data are available on the French open data plateform recherche.data.gouv at \url{http://dx.doi.org/10.57745/FJFLYB}.
	
	\section*{Acknowledgments}
	The authors gratefully thank Jean-Baptiste Leblond for stimulating discussions on the second-order derivation, Antoine Sanner and Lars Pastewka for fruitful discussions on the extension of our approach to other physical systems and on variational formulations, Sébastien Brisard for suggesting variance reduction methods. We also thank the anonymous reviewer for their constructive and insightful comments. ML thanks Antoine Sanner for a critical reading of the manuscript.\\
	
	For the purpose of Open Access, a CC-BY public copyright licence has been applied by the authors to the present document and will be applied to all subsequent versions up to the Author Accepted Manuscript arising from this submission.\\
	
	\noindent \includegraphics[width=3cm]{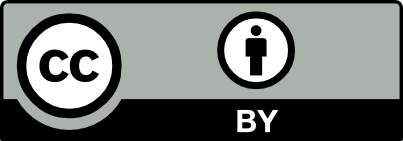} \href{https://creativecommons.
		org/licenses/by/4.0/}{\textbf{Distributed under a Creative Commons Attribution | 4.0 International licence}}
	
	\appendix
	
	\section{Calculation of the first-order variations of the kernel $\delta Z$}
	\label{apdx:calculation_Z}
	
	In this Appendix, our aim is to provide useful indications on the derivation of Eq.~\eqref{eq:perturbated_kernel_1stOrder} that gives the first-order variations $\delta Z$ of the kernel, starting from the general formula~\eqref{eq:perturbated_kernel_general}. Following Eqs.~\eqref{eq:kernel_circular} and \eqref{eq:kernel_star_variation}, the formula~ \eqref{eq:perturbated_kernel_general} writes as:
	\begin{equation}
		\label{eq:kernel_calculation_01}
		\scalemath{0.9}{
			\begin{aligned}
				\delta Z(\theta_1, \theta_2) & = Z^0(a_0; \theta_1, \theta_2) \left[\left(\dfrac{\mathcal{D}(M_1, M_2)}{\mathcal{D}(M^0_1, M^0_2)}\right)^{-2}-1\right] + \mathrm{PV}\int_{0}^{2\pi} Z^0(a_0; \theta_1, \theta) Z^0(a_0; \theta_2, \theta) \left[\delta a(\theta) - \delta a_*(\theta)\right] a_0 d\theta + \mathcal{O}(A^2) \\
				& = -\dfrac{\delta a(\theta_1) + \delta a(\theta_2)}{8\pi a_0^3\sin^2\left[(\theta_2-\theta_1)/2\right]} + \dfrac{1}{64\pi^2a_0^3}\mathrm{PV}\int_{0}^{2\pi} 	\dfrac{1}{\sin^2[(\theta-\theta_1)/2]\sin^2[(\theta-\theta_2)/2]} \left[\delta a(\theta) - \delta a_*(\theta)\right] a_0 d\theta + \mathcal{O}(A^2).
			\end{aligned}
		}
	\end{equation}
	Next, we use the first-order expansion of $\delta a_*$ of Eq.~\eqref{eq:star_transformation_1stOrder}, and the decomposition of the sinus fraction given in Eq.~\eqref{eq:partial_fraction_decomposition}:
	\begin{equation}
		\label{eq:kernel_calculation_02}
		\scalemath{0.85}{
			\begin{aligned}
				\delta Z(\theta_1, \theta_2) = & -\dfrac{\delta a(\theta_1) + \delta a(\theta_2)}{8\pi a_0^3\sin^2\left[(\theta_2-\theta_1)/2\right]} + \dfrac{1}{(8\pi)^2a_0^3\sin^2\left[(\theta_2-\theta_1)/2\right]} \times \left\lbrace \rule{0cm}{1cm}\right. \\
				& \mathrm{PV}\int_{0}^{2\pi} \dfrac{\cos^2[(\theta-\theta_1)/2]}{\sin^2[(\theta-\theta_1)/2]} \left[\delta a(\theta) - \delta a(\theta_1)\right] d\theta + \mathrm{PV}\int_{0}^{2\pi} \dfrac{\cos^2[(\theta-\theta_2)/2]}{\sin^2[(\theta-\theta_2)/2]} \left[\delta a(\theta) - \delta a(\theta_2)\right] d\theta \\
				& - \dfrac{\delta a(\theta_2) - \delta a(\theta_1)}{2} \cdot \dfrac{\cos[(\theta_2-\theta_1)/2]}{\sin[(\theta_2-\theta_1)/2]}  \cdot \left( \mathrm{PV}\int_{0}^{2\pi} \underbrace{\dfrac{\cos^2[(\theta-\theta_1)/2]}{\sin^2[(\theta-\theta_1)/2]} \sin(\theta-\theta_1)}_\text{odd integrand} d\theta + \mathrm{PV}\int_{0}^{2\pi} \underbrace{\dfrac{\cos^2[(\theta-\theta_2)/2]}{\sin^2[(\theta-\theta_2)/2]} \sin(\theta-\theta_2)}_\text{odd integrand} d\theta \right) \\
				& - \dfrac{\delta a(\theta_2) - \delta a(\theta_1)}{2} \cdot \left( \underbrace{\int_{0}^{2\pi} \cos^2[(\theta-\theta_1)/2] d\theta - \int_{0}^{2\pi} \cos^2[(\theta-\theta_2)/2] d\theta}_\text{cancel out} \right) \\
				& + \dfrac{2\cos[(\theta_2-\theta_1)/2]}{\sin[(\theta_2-\theta_1)/2]} \cdot \mathrm{PV}\int_{0}^{2\pi} \left[ \dfrac{\cos^3[(\theta-\theta_1)/2]}{\sin[(\theta-\theta_1)/2]} - \dfrac{\cos^3[(\theta-\theta_2)/2]}{\sin[(\theta-\theta_2)/2]} \right] \delta a(\theta) d\theta \\
				& - \left(\mathrm{PV}\int_{0}^{2\pi} \underbrace{\dfrac{\cos^3[(\theta-\theta_1)/2]}{\sin[(\theta-\theta_1)/2]}}_\text{odd integrand} d\theta\right) \delta a(\theta_1) - \left( \mathrm{PV}\int_{0}^{2\pi} \underbrace{\dfrac{\cos^3[(\theta-\theta_2)/2]}{\sin[(\theta-\theta_2)/2]}}_\text{odd integrand} d\theta \right) \delta a(\theta_2) \\
				& - \dfrac{\delta a(\theta_2) - \delta a(\theta_1)}{2} \cdot \dfrac{2\cos^2[(\theta_2-\theta_1)/2]}{\sin^2[(\theta_2-\theta_1)/2]} \cdot \left(\underbrace{\int_{0}^{2\pi} \dfrac{\cos^3[(\theta-\theta_1)/2]}{\sin[(\theta-\theta_1)/2]} \sin(\theta-\theta_1) d\theta - \int_{0}^{2\pi} \dfrac{\cos^3[(\theta-\theta_2)/2]}{\sin[(\theta-\theta_2)/2]} \sin(\theta-\theta_2) d\theta}_\text{cancel out} \right) \\
				& - \dfrac{\delta a(\theta_2) - \delta a(\theta_1)}{2} \cdot \dfrac{2\cos[(\theta_2-\theta_1)/2]}{\sin[(\theta_2-\theta_1)/2]} \cdot \left( \int_{0}^{2\pi} \underbrace{\dfrac{\cos^3[(\theta-\theta_1)/2]}{\sin[(\theta-\theta_1)/2]} 2\sin^2[(\theta-\theta_1)/2]}_\text{odd integrand} d\theta + \int_{0}^{2\pi} \underbrace{\dfrac{\cos^3[(\theta-\theta_2)/2]}{\sin[(\theta-\theta_2)/2]} 2\sin^2[(\theta-\theta_2)/2]}_\text{odd integrand} d\theta \right) \\
				& + \int_{0}^{2\pi} 2\cos^2[(\theta-\theta_1)/2] \left[ \delta a(\theta) - \delta a(\theta_1) \right] d\theta + \int_{0}^{2\pi} 2\cos^2[(\theta-\theta_2)/2] \left[ \delta a(\theta) - \delta a(\theta_2) \right] d\theta \\
				& - \dfrac{\delta a(\theta_2) - \delta a(\theta_1)}{2} \cdot \dfrac{\cos[(\theta_2-\theta_1)/2]}{\sin[(\theta_2-\theta_1)/2]} \left(\int_{0}^{2\pi} 2\underbrace{\cos^2[(\theta-\theta_1)/2] \sin(\theta-\theta_1)}_\text{odd integrand} d\theta + \int_{0}^{2\pi} 2\underbrace{\cos^2[(\theta-\theta_2)/2] \sin(\theta-\theta_2)}_\text{odd integrand} d\theta \right) \\
				& \left. - \dfrac{\delta a(\theta_2) - \delta a(\theta_1)}{2} \cdot \left(\underbrace{\int_{0}^{2\pi} 4\cos^2[(\theta-\theta_1)/2] \sin^2[(\theta-\theta_1)/2] d\theta - \int_{0}^{2\pi} 4\cos^2[(\theta-\theta_2)/2] \sin^2[(\theta-\theta_2)/2] d\theta}_\text{cancel out}\right) \right\rbrace + \mathcal{O}(A^2).
			\end{aligned}
		}
	\end{equation}
	After simplifications of the integrals that are zero due to the oddness of the integrand, and those that cancel out, one has:
	\begin{equation}
		\label{eq:kernel_calculation_03}
		\scalemath{0.95}{
			\begin{aligned}
				\delta Z(\theta_1, \theta_2) = & -\dfrac{\delta a(\theta_1) + \delta a(\theta_2)}{8\pi a_0^3\sin^2\left[(\theta_2-\theta_1)/2\right]} + \dfrac{1}{(8\pi)^2a_0^3\sin^2\left[(\theta_2-\theta_1)/2\right]} \times \Bigg\lbrace \\
				& \hspace{15mm} \mathrm{PV}\int_{0}^{2\pi} \dfrac{\cos^2[(\theta-\theta_1)/2]}{\sin^2[(\theta-\theta_1)/2]} \left[\delta a(\theta) - \delta a(\theta_1)\right] d\theta + \mathrm{PV}\int_{0}^{2\pi} \dfrac{\cos^2[(\theta-\theta_2)/2]}{\sin^2[(\theta-\theta_2)/2]} \left[\delta a(\theta) - \delta a(\theta_2)\right] d\theta \\
				& \hspace{15mm} + \dfrac{2\cos[(\theta_2-\theta_1)/2]}{\sin[(\theta_2-\theta_1)/2]} \cdot \mathrm{PV}\int_{0}^{2\pi} \left[ \dfrac{\cos^3[(\theta-\theta_1)/2]}{\sin[(\theta-\theta_1)/2]} - \dfrac{\cos^3[(\theta-\theta_2)/2]}{\sin[(\theta-\theta_2)/2]} \right] \delta a(\theta) d\theta \\
				& \hspace{15mm} \left. + \int_{0}^{2\pi} 2\left[\cos^2\left(\dfrac{\theta-\theta_1}{2}\right) + \cos^2\left(\dfrac{\theta-\theta_2}{2}\right)\right] \delta a(\theta) d\theta - 2\pi\left[\delta a(\theta_1) + \delta a(\theta_2)\right] \right\rbrace + \mathcal{O}(A^2),
			\end{aligned}
		}
	\end{equation}
	from which one retrieves Eq.~\eqref{eq:perturbated_kernel_1stOrder}. Note that the simplifications in Eq.~\eqref{eq:kernel_calculation_02} arise from the careful choice of $\delta a_*$ in Eq.~\eqref{eq:star_transformation_general} and that of $\theta_*$ in Eq.~\eqref{eq:star_transformation_theta_star}, which allows for the alternate expressions of $\delta a_*$ in Eq.~\eqref{eq:star_transformation_1stOrder} that pair well with the decomposition of Eq.~\eqref{eq:partial_fraction_decomposition}. 
	
	\section{Calculation of the second-order variations of the SIF $\delta K$}
	\label{apdx:calculation_K}
	
	In this Appendix, we provide useful indications on how to calculate the second-order $\KI^2$ perturbations of the mode I stress intensity factor arising from perturbations $\delta a = A\phi$ of the circular front of radius $a_0$. For the sake of completeness, we do it here in the general case where the reference SIF $\KI^0$, applied to the unperturbed circular crack, depends on $\theta$. The calculations presented in the main text only address cases where $\KI^0$ is axisymmetric.\\
	
	As established in Eq.~\eqref{eq:SIF_variation_02}, the SIF variations arising from an additional infinitesimal perturbation $\delta A\phi$ reads:
	\begin{equation}
		\label{eq:SIF_calculation_01}
		\modif{
		\begin{aligned}
			& \KI(a_0, \left[(A+\delta A)\phi\right]; \theta_1) - \KI(a_0, \left[A\phi\right]; \theta_1) =  \KI(a_0+\delta A\phi(\theta_1), \left[A\phi\right]; a_0; \theta_1) - \KI(a_0, \left[A\phi\right]; \theta_1) \\ 
			& \hspace{2cm} + \mathrm{PV}\int_{0}^{2\pi} Z(a_0+\delta A\phi(\theta_1), \left[A\phi\right]; \theta_1, \theta) \KI(a_0+\delta A\phi(\theta_1), \left[A\phi\right];\theta) \left[\delta A\phi(\theta) - \delta A\phi(\theta_1)\right] ds(\theta) + \mathcal{O}(\delta A^2),
		\end{aligned}
		}
	\end{equation}
	where $s$ is the curvilinear abscissa along the perturbed front \modif{$\mathcal{F'}$}, so that $ds$ reduces here to:
	\begin{equation}
		\label{eq:curvilinear_abscissa}
		\modif{
		ds(\theta) = a_0 + \delta A\phi(\theta_1) + A\phi(\theta) + \mathcal{O}(A^2),
		}
	\end{equation}
	Using Eq.~\eqref{eq:perturbed_SIF_1stOrder}, one has:
	\begin{equation}
		\label{eq:SIF_calculation_02}
		\modif{
		\begin{aligned}
			& \KI(a_0+\delta A\phi(\theta_1), \left[A\phi\right]; a_0; \theta_1) - \KI(a_0, \left[A\phi\right]; \theta_1) \\
			& \hspace{2cm} = \delta A \times \left\lbrace \dfrac{\partial \KI^0}{\partial a}(a_0, \theta_1)\phi(\theta_1) + A\dfrac{\partial^2\KI^0}{\partial a^2}(a_0, \theta_1)\phi(\theta_1)^2 \right. \\
			& \hspace{3cm} + A\, \mathrm{PV}\int_{0}^{2\pi} \dfrac{\partial Z^0}{\partial a}(a_0; \theta_1, \theta) \KI^0(a_0, \theta) \left[\phi(\theta) - \phi(\theta_1)\right] \phi(\theta_1) a_0 d\theta \\
			& \hspace{3cm} + A\, \mathrm{PV}\int_{0}^{2\pi} Z^0(a_0; \theta_1, \theta) \dfrac{\partial \KI^0}{\partial a}(a_0, \theta) \left[\phi(\theta) - \phi(\theta_1)\right] \phi(\theta_1) a_0 d\theta \\
			& \hspace{3cm} \left. + A\, \mathrm{PV}\int_{0}^{2\pi} Z^0(a_0; \theta_1, \theta) \KI^0(a_0, \theta) \left[\phi(\theta) - \phi(\theta_1)\right] \phi(\theta_1) d\theta + \mathcal{O}(A^2) \right\rbrace + \mathcal{O}(\delta A^2).
		\end{aligned}
		}
	\end{equation}
	The calculation of the second term in the right-hand side of Eq.~\eqref{eq:SIF_calculation_01} is also straightforward. From Eqs.~\eqref{eq:SIF_2ndOrder_general} and \eqref{eq:curvilinear_abscissa}, one has:
	\begin{equation}
		\label{eq:SIF_calculation_03}
		\modif{
		\begin{aligned}
			\mathrm{PV}\int_{0}^{2\pi} & Z(a_0+\delta A\phi(\theta_1), \left[A\phi\right]; \theta_1, \theta) \KI(a_0, \left[A\phi\right];\theta) \left[\delta A\phi(\theta) - \delta A\phi(\theta_1)\right] ds(\theta) \\
			= \mathrm{PV}\int_{0}^{2\pi} & Z(a_0, \left[A\phi\right]; \theta_1, \theta) \KI(A; a_0; \theta) \left[\delta A\phi(\theta) - \delta A\phi(\theta_1)\right] ds(\theta)  + \mathcal{O}(\delta A^2) \\
			& = \delta A \times \left\lbrace \mathrm{PV}\int_{0}^{2\pi} Z^0(a_0; \theta_1, \theta) \KI^0(a_0, \theta) \left[\phi(\theta) - \phi(\theta_1)\right] a_0 d\theta \right. \\
			& \hspace{6mm} + A\, \mathrm{PV}\int_{0}^{2\pi} Z^1(a_0, \phi; \theta_1, \theta) \KI^0(a_0, \theta) \left[\phi(\theta) - \phi(\theta_1)\right] a_0 d\theta \\
			& \hspace{6mm} + A\, \mathrm{PV}\int_{0}^{2\pi} Z^0(a_0; \theta_1, \theta) \dfrac{\partial \KI^0}{\partial a}(a_0, \theta) \left[\phi(\theta) - \phi(\theta_1)\right] \phi(\theta) a_0 d\theta \\
			& \hspace{6mm} + A\, \mathrm{PV}\int_{0}^{2\pi} \int_{0}^{2\pi} Z^0(a_0; \theta_1, \theta) Z^0(a_0; \theta, \theta') \KI^0(a_0, \theta') \left[\phi(\theta) - \phi(\theta_1)\right] \left[\phi(\theta') - \phi(\theta)\right] a_0^2 d\theta d\theta' \\
			& \hspace{6mm} \left. + A\, \mathrm{PV}\int_{0}^{2\pi} Z^0(a_0; \theta_1, \theta) \KI^0(a_0, \theta) \left[\phi(\theta) - \phi(\theta_1)\right] \phi(\theta) d\theta + \mathcal{O}(A^2) \right\rbrace + \mathcal{O}(\delta A^2).
		\end{aligned}
		}
	\end{equation}
	Identifying the terms in $\delta A$ in the expansions of Eqs.~\eqref{eq:SIF_variation_01} and \eqref{eq:SIF_calculation_01}, one finds:
	\begin{equation}
		\label{eq:SIF_calculation_04}
		\KI^1(a_0, [\KI^0], [\phi]; \theta_1) = \dfrac{\partial \KI^0}{\partial a}(a_0, \theta_1)\phi(\theta_1) + \mathrm{PV}\int_{0}^{2\pi} Z^0(a_0; \theta_1, \theta) \KI^0(a_0, \theta) \left[\phi(\theta) - \phi(\theta_1)\right] a_0 d\theta,
	\end{equation}
	which is Eq.~\eqref{eq:perturbed_SIF_1stOrder} of \cite{gao_somewhat_1987}. The terms in $\delta A \times A$ yield:
	\begin{equation}
		\label{eq:SIF_calculation_05}
		\begin{aligned}
			2\KI^2(a_0, [\KI^0], [\phi]; \theta_1) = & \dfrac{\partial^2\KI^0}{\partial a^2}(a_0, \theta_1)\phi(\theta_1)^2 + \mathrm{PV}\int_{0}^{2\pi} Z^0(a_0; \theta_1, \theta) \dfrac{\partial \KI^0}{\partial a}(a_0, \theta) \left[\phi(\theta)^2 - \phi(\theta_1)^2\right] a_0 d\theta\\
			& + \mathrm{PV}\int_{0}^{2\pi} Z^0(a_0; \theta_1, \theta) \KI^0(a_0, \theta) \left[\phi(\theta) - \phi(\theta_1)\right]^2 d\theta \\
			& + \mathrm{PV}\int_{0}^{2\pi} \int_{0}^{2\pi} Z^0(a_0; \theta_1, \theta) Z^0(a_0; \theta, \theta') \KI^0(a_0, \theta') \left[\phi(\theta) - \phi(\theta_1)\right] \left[\phi(\theta) - \phi(\theta')\right] a_0^2 d\theta d\theta' \\
			& + \mathrm{PV}\int_{0}^{2\pi} Z^1(a_0, \left[\phi\right]; \theta_1, \theta) \KI^0(a_0, \theta) \left[\phi(\theta) - \phi(\theta_1)\right] a_0 d\theta,
		\end{aligned}
	\end{equation}
	which is Eq.~\eqref{eq:perturbed_SIF_K2}. Here we used $\partial Z^0 / \partial a = -2Z^0/a_0$ from Eq.~\eqref{eq:kernel_circular}.\\
	
	We need now to calculate Eqs.~\eqref{eq:SIF_calculation_04} and \eqref{eq:SIF_calculation_05} by resorting to Fourier series. The calculations are heavy, and we will show them here only for $\KI^1$. Eq.~\eqref{eq:SIF_calculation_04} yields:
	\begin{equation}
		\label{eq:SIF_calculation_06}
		\scalemath{0.95}{
			\begin{aligned}
				\KI^1(a_0, [\KI^0], [\phi]; \theta_1) & = \dfrac{\partial \KI^0}{\partial a}(a_0, \theta_1)\phi(\theta_1) + \mathrm{PV}\int_{0}^{2\pi} Z^0(a_0; \theta_1, \theta) \KI^0(a_0, \theta) \left[\phi(\theta) - \phi(\theta_1)\right] a_0 d\theta \\
				& = \sum_{k, k'} \widehat{\left(\dfrac{\partial\KI^0}{\partial a}\right)}_k\!(a_0) \, \hat{\phi}_{k'} e^{i(k+k')\theta_1} + \sum_{k, k'} \widehat{K_{\mathrm{I}k}^0}(a_0) \hat{\phi}_{k'} \left[\dfrac{1}{8a_0} \mathrm{PV}\int_{0}^{2\pi} \dfrac{e^{ik'(\theta-\theta_1)}-1}{\sin^2[(\theta-\theta_1)/2]} e^{ik(\theta-\theta_1)} d\theta\right] e^{i(k+k')\theta_1} \\
				& = \sum_{k, k'} \left[ \widehat{\left(\dfrac{\partial\KI^0}{\partial a}\right)}_k\!(a_0) - \dfrac{\abs{k+k'}-\abs{k}}{2a_0} \widehat{K_{\mathrm{I}k}^0}(a_0) \right] \hat{\phi}_{k'} e^{i(k+k')\theta_1},
			\end{aligned}	
		}
	\end{equation}
	where $\hat{\phi}_k$, $\widehat{K_{\mathrm{I}k}^0}(a_0)$, and $\widehat{\left(\dfrac{\partial\KI^0}{\partial a}\right)}_k\!(a_0)$ are the $k$-th Fourier coefficient of $\phi$, $\KI^0(a_0, \cdot)$, and $\dfrac{\partial\KI}{\partial a}(a_0, \cdot)$ respectively. The terms into brack in Eq.~\eqref{eq:SIF_calculation_06} reduces to $L(k)$ as given in Eq.~(\ref{eq:fourier_kernels_2ndOrder}a) when $\KI^0$ is independent of $\theta$. In deriving \eqref{eq:SIF_calculation_06}, we used the following integral:
	\begin{equation}
		\label{eq:useful_integrals_01}
		\mathrm{PV}\int_{0}^{2\pi} \dfrac{e^{iku}-1}{\sin^2(u/2)} du = -4\pi\abs{k},
	\end{equation}
	which is calculated using the residue theorem. The calculations for $\KI^2$ are similar, although more complicated as they involve triple sums. They can be simplified by building on the following expression of the first-order kernel variations $Z^1$:
	\begin{equation}
		\label{eq:fourier_kernel_Z1}
		\scalemath{0.9}{
			\begin{aligned}
				Z^1(a_0, \left[\phi\right]; \theta_1, \theta_2) = - \dfrac{1}{16\pi a_0^2\sin^2\left[(\theta_2-\theta_1)/2\right]} \times \left[ \rule{0cm}{0.6cm} \right. & \sum_k \hat{\phi}_k \left(e^{ik\theta_2} + e^{ik\theta_1}\right) \left(\abs{k} +2 - \dfrac{\Delta_{\abs{k}1}}{4}\right) \\
				& + \left. i\,\dfrac{\cos\left[(\theta_2-\theta_1)/2\right]}{\sin\left[(\theta_2-\theta_1)/2\right]}\sum_k \hat{\phi}_k \left(e^{ik\theta_2} - e^{ik\theta_1}\right) \left(1 - \Delta_{k0} - \dfrac{\Delta_{\abs{k}1}}{4} \right) \mathrm{sign}(k) \right],
			\end{aligned}
		}
	\end{equation}
	as well as the following integrals:
	\begin{equation}
		\label{eq:useful_integrals_02}
		\begin{aligned}
			& \mathrm{PV}\int_{0}^{2\pi} \dfrac{2\cos^3(u/2)}{\sin(u/2)} e^{iku} du = 4i\pi\,\mathrm{sign}(k)\left(1-\dfrac{1}{4}\Delta_{\abs{k}1}-\Delta_{k0}\right) \\
			& \mathrm{PV}\int_{0}^{2\pi} \dfrac{\cos(u/2)}{\sin^3(u/2)}(e^{iku}-1)(e^{ik'u}-1) du = -4i\pi\left[k(\abs{k+k'}-\abs{k})+k'(\abs{k+k'}-\abs{k'})\right],
		\end{aligned}
	\end{equation}
	where $\Delta_{ij}$ is the Kronecker symbol that is equal to $1$ when $i=j$ and $0$ otherwise, and $\mathrm{sign}(k)$ is the sign of the integer $k$. Using Eqs.~\eqref{eq:useful_integrals_01}, \eqref{eq:fourier_kernel_Z1}, and \eqref{eq:useful_integrals_02}, Eq.~\eqref{eq:SIF_calculation_05} yields:
	\begin{equation}
		\label{eq:SIF_calculation_07}
		\KI^2(a_0, [\KI^0], [\phi]; \theta_1) = \sum_{k,k',k''} \mathcal{H}\left(\left[\KI^0\right]; k,k',k''\right) \hat{\phi}_{k'} \hat{\phi}_{k''} e^{i(k+k'+k'')\theta_1},
	\end{equation}
	where the convolution kernel $\mathcal{K}$ reads:
	\begin{equation}
		\label{eq:SIF_calculation_08}
		\scalemath{0.95}{
			\begin{aligned}
				\mathcal{H}\left(\left[\KI^0\right]; k,k',k''\right) = & \dfrac{1}{2} \widehat{\left(\dfrac{\partial^2\KI^0}{\partial a^2}\right)}_k\!(a_0) - \dfrac{1}{4a_0} \widehat{\left(\dfrac{\partial\KI^0}{\partial a}\right)}_k\!(a_0) \left(\abs{k+k'+k''}-\abs{k}\right)
				- \dfrac{1}{4a_0^2}\widehat{K_{\mathrm{I}k}^0}(a_0) \left( \abs{k+k'+k''} - \abs{k+k'} - \abs{k+k''} + \abs{k} \right) \\
				& - \dfrac{1}{16a_0^2}\widehat{K_{\mathrm{I}k}^0}(a_0) \left( (k+k')^2 + (k+k'')^2 \right)
				+ \dfrac{1}{16a_0^2}\widehat{K_{\mathrm{I}k}^0}(a_0) \left( \abs{k+k'} + \abs{k+k''} \right) \left( \abs{k+k'+k''} + \abs{k} \right) \\
				& - \dfrac{1}{8a_0^2}\widehat{K_{\mathrm{I}k}^0}(a_0) \abs{k} \abs{k+k'+k''}
				+ \dfrac{1}{4a_0^2}\widehat{K_{\mathrm{I}k}^0}(a_0) \left( \abs{k+k'+k''} - \abs{k} \right)
				+ \dfrac{1}{16a_0^2}\widehat{K_{\mathrm{I}k}^0}(a_0) \left( \abs{k'}\abs{k+k''} + \abs{k''}\abs{k+k'} \right) \\
				& - \dfrac{1}{16a_0^2}\widehat{K_{\mathrm{I}k}^0}(a_0) \abs{k} \left( \abs{k'} + \abs{k''} \right)
				- \dfrac{1}{16a_0^2}\widehat{K_{\mathrm{I}k}^0}(a_0) k \left( \abs{k+k'+k''} - \abs{k+k'} - \abs{k+k''} + \abs{k} \right) \left( \mathrm{sign}(k') + \mathrm{sign}(k'')\right) \\
				& - \dfrac{1}{16a_0^2}\widehat{K_{\mathrm{I}k}^0}(a_0) \left( \abs{k+k'+k''}(\abs{k'} + \abs{k''}) - \abs{k'}\abs{k+k'} - \abs{k''}\abs{k+k''} \right) \mathrm{sign}(k'k'').
			\end{aligned}	
		}
	\end{equation}
	The set of equations~(B.11-12) is validated next in \ref{apdx:validation_K} on elliptical cracks loaded by a non-axisymmetric loading. Considering the limiting case of a reference SIF $\KI^0$ independent of $\theta$, one has:
	\begin{equation}
		\label{eq:SIF_calculation_09}
		\begin{aligned}
			\mathcal{H}\left(\left[\KI^0\right]; k,k',k''\right) & = \dfrac{1}{2} \dfrac{d^2\KI}{d a^2}\!(a_0) - \dfrac{1}{a_0} \dfrac{d\KI}{da}(a_0) \dfrac{\abs{k+k'}}{4}
			+ \dfrac{1}{a_0^2} \KI^0(a_0) \dfrac{\abs{k} + \abs{k'}}{4} \\
			& + \dfrac{1}{8a_0^2}\KI^0(a_0)\abs{k}\abs{k'}
			+ \dfrac{1}{16a_0^2}\KI^0(a_0) \left( \abs{k+k'}(\abs{k} + \abs{k'}) - k^2 - {k'}^2 \right) \left(1 - \mathrm{sign}(kk')\right) \\
			& = \dfrac{1}{2} \dfrac{d^2\KI}{d a^2}\!(a_0) - \dfrac{1}{a_0} \dfrac{d\KI}{da}(a_0) \dfrac{\abs{k+k'}}{4} + \dfrac{1}{a_0^2}\KI^0(a_0) \dfrac{\abs{k} + \abs{k'}}{4} + \dfrac{1}{16a_0^2}\KI^0(a_0) R(k,k'),
		\end{aligned}	
	\end{equation}
	from which one finds back Eq.~\eqref{eq:perturbed_SIF_K2}.\\
	
	The general expression~(B.11-12) of the second-order variations of stress intensity factor for a perturbed crack front given prove rather complex, and its scope of application seems limited. Indeed, from a theoretical point of view, the influence of a non-uniformity of macroscopic loading on mode I crack propagation is already of second-order within the linear expansion of \cite{gao_somewhat_1987} (see Eq.~\eqref{eq:perturbed_SIF_1stOrder}). Moreover, the triple sum over the Fourier coefficients related to $\KI^0$, its derivatives with crack advance $(\partial^n \KI^0/\partial a^n)$, and the perturbation shape $\phi$ makes it unfit for numerical calculations, as estimating second-order contributions would take a computation time in $\mathcal{O}(n^3\ln n)$, where $n$ is the number of discretization points along the crack front. This is already larger than more conventional numerical approaches, like e.g. the boundary element method, that can simulate crack propagation \emph{at all orders} in ``only'' $\mathcal{O}(n^2\ln n)$.
	
	However, we did not perform here the rigorous calculation for a general $\KI^0$ for the sake of completeness only. Indeed, for shear ruptures, the mode II and mode III stress intensity factors applied to a circular crack depends on $\theta$, even when the applied stress is axisymmetric \citep{gao_nearly_1988}. Propagating internal shear cracks often adopt quasi-elliptical front shapes \citep{saez_three-dimensional_2022} that may be described with a perturbative approach to LEFM. We provide thus in this Appendix the basic steps needed to derive an equivalent second-order theory for shear cracks. Future work will be devoted to the construction of this model and its application to the propagation of e.g. fluid-driven shear ruptures.
	
	\section{Validation of the general formula \eqref{eq:SIF_calculation_07} on elliptical cracks loaded by non-axisymmetric loadings}
	\label{apdx:validation_K}
	
	This Appendix aims at validating the general expression of the second-order variations of stress intensity factor along a perturbed crack front obtained in Eqs.~(B.11-12). We provided in Section~3.1 a validation of Eq.~\eqref{eq:perturbed_SIF_K2} or equivalently \eqref{eq:SIF_calculation_09} by comparing our second-order predictions to the second-order expansions of the exact stress intensity factor distribution along elliptical crack fronts loaded by a uniform remote stress $\sigma$. Fortunately, analytical predictions for the SIF distribution along elliptical crack fronts are available in the literature for more complex loading situations \citep{shah_stress_1971, kassir_3D_1975, atroshchenko_stress_2009}. Here, a comparison between our model's predictions and second-order expansions of exact analytical solutions is provided for three non-axisymmetric loading cases.\\
	
	As in Section 3.1, we consider a crack whose front $\mathcal{F}$ describes an ellipsis centered in $O$, of short semi-axis $c$ aligned with $\mathbf{e_z}$, and long semi-axis $b$ aligned with $\mathbf{e_x}$. The crack front $\mathcal{F}$ follows the contour parametrized by the variable $\varphi$: $\left(z = b\cos\varphi, x=c\sin\varphi\right)$. Its position $a(\theta)$ in polar coordinates is given by Eq.~\eqref{eq:ellipsis_transformation}, \review{so that:}
\begin{equation}
		\review{a(\theta)\cos\theta = b\cos\varphi \text{, and } a(\theta)\sin\theta = c\sin\varphi.}
	\end{equation}
	
We follow here the reasoning of \cite{gao_somewhat_1987}, and take as small expansion parameter $\epsilon = 1-c/b$, and a reference radius $a_0$ equal to the front extension at the point $\theta_1$ of SIF evaluation (i.e. $a_0 = a(\theta_1)$). In that case, the perturbation $\delta a$ reads:
	\begin{equation}
		\label{eq:validation_local_perturbation}
		\delta a(\theta) = a_0 \left(1 + \epsilon A_1(\theta) + \epsilon^2 A_2(\theta)\right),
	\end{equation}
	where
	\begin{equation}
		\label{eq:validation_local_shapes}
		\begin{aligned}
			& A_1(\theta) = -\dfrac{1}{2}\cos(2\theta_1) + \dfrac{1}{2} \cos(2\theta),\\
			\text{and } &A_2(\theta) = \dfrac{1}{8} - \dfrac{1}{4}\cos(2\theta_1) - \dfrac{1}{16}\cos(4\theta_1) + \left( \dfrac{1}{4} - \dfrac{1}{4}\cos(2\theta_1)\right)\cos(2\theta) + \dfrac{3}{16}\cos(4\theta).
		\end{aligned}
	\end{equation}
	
	Next, we introduce $k_\mathrm{I}$, the crack face weight function of a circular crack. $k_\mathrm{I}(a, \theta; r, \theta')$ corresponds to the mode I SIF generated at the point $\theta$ along the circular front of radius $a$ by a pair of unitary symmetric tensile forces acting on the point interior to the crack indexed by $(r, \theta')$. It reads \citep{tada_stress_2000}:
	\begin{equation}
		\label{eq:crackface_weightfunction}
		k_\mathrm{I}(a, \theta; r, \theta') = \dfrac{\sqrt{a^2-r^2}}{\pi^{3/2}a^{1/2} \left[a^2+r^2-2ar\cos(\theta-\theta')\right]}.
	\end{equation}
	Considering that the crack is loaded by a distribution of tensile stress
	\begin{equation*}
		\sigma(r, \theta) = \sum_k \hat{\sigma}_k(r) e^{ik\theta},
	\end{equation*}
	acting on the failure plane $y=0$ of the intact body, the mode I SIF generated along the front of a circular crack of radius $a$ reads:
	\begin{equation}
		\label{eq:validation_reference_SIF}
		\begin{aligned}
			\KI^0(a,\theta) & = \int_{0}^{2\pi} \int_{0}^{a} \sigma(r, \theta') k_\mathrm{I}(a, \theta; r, \theta') rdrd\theta'\\
			& = \sum_k \left[\dfrac{2}{\pi} \sqrt{\pi a} \int_0^1 \dfrac{u^{\abs{k}+1}}{\sqrt{1-u^2}} \hat{\sigma}_k(au) du\right] e^{ik\theta}.
		\end{aligned}
	\end{equation}
	We have at our disposal all the basic ingredients to use our second-order model of Eqs.~\eqref{eq:SIF_calculation_06} and \eqref{eq:SIF_calculation_07}.
	
	\paragraph{Linear stress field dependent on $z$}~\\
	
	We first consider the case where the elliptical crack is loaded by a linear stress field dependent on $z$:
	\begin{equation}
		\label{eq:validation_cos_sigma}
		\sigma(z,x) = \sigma_0 \dfrac{z}{b} \Rightarrow \sigma(r,\theta) = \sigma_0 \dfrac{r}{b}\cos(\theta).
	\end{equation}
	Using Eq.~\eqref{eq:validation_reference_SIF} the associated SIF distribution along the front of a circular crack of radius $a_0$ reads:
	\begin{equation}
		\label{eq:validation_cos_KI0}
		\KI^0(a,\theta) = \dfrac{4}{3\sqrt{\pi}}\dfrac{a^{3/2}}{b} \sigma_0 \cos(\theta).
	\end{equation}
	The exact SIF distribution along the elliptical crack front has been calculated by \citep{shah_stress_1971} and \citep{atroshchenko_stress_2009}. It reads:
	\begin{equation}
		\label{eq:validation_cos_KIe}
		\begin{aligned}
			&\KI^e(\varphi) = - \sqrt{\pi c} \sigma_0\cos(\varphi) \dfrac{\kappa^2}{(1-2\kappa^2)E(\kappa) - {\kappa'}^2K(\kappa)} \left((c/b)^4\cos^2(\varphi)+\sin^2(\varphi)\right)^{1/4} \\
			\Rightarrow\, & \KI^e(\theta) = - \dfrac{\sqrt{\pi}c^{3/2}}{b} \sigma_0 \cos(\theta) \dfrac{\kappa^2}{(1-2\kappa^2)E(\kappa) - {\kappa'}^2K(\kappa)} \dfrac{\left((c/b)^4\cos^2(\theta)+\sin^2(\theta)\right)^{1/4}}{\left((c/b)^2\cos^2(\theta)+\sin^2(\theta)\right)^{3/4}}.\\
		\end{aligned}	
	\end{equation}
	where $\kappa = \left(1-c^2/b^2\right)^{1/2}$ and $\kappa' = c/b$. 
	
	One may then use Eqs.~\eqref{eq:validation_cos_KI0} and \eqref{eq:validation_cos_KIe} to expand $\KI^e/\KI^0$ to second order in $\epsilon$ at $\epsilon=0$ and $\theta = \theta_1$. One finds:
	\begin{equation}
		\label{eq:validation_cos_expansion}
		\begin{aligned}
			\KI^e(A_1)/\KI^0(a_0,\theta_1) & = -\dfrac{3\pi}{4} \dfrac{\kappa^2}{(1-2\kappa^2)E(\kappa) - {\kappa'}^2K(\kappa)} \left((c/b)^4\cos^2(\theta_1)+\sin^2(\theta_1)\right)^{1/4} \\
			& = 1 + \epsilon \left[\dfrac{1}{4} - \dfrac{1}{2}\cos(2\theta_1)\right] + \epsilon^2 \left[\dfrac{5}{16} - \dfrac{3}{8}\cos(2\theta_1) - \dfrac{3}{16}\cos(4\theta_1)\right] + \mathcal{O}(\epsilon^3).
		\end{aligned}	
	\end{equation}
	Our aim here is to retrieve Eq.~\eqref{eq:validation_cos_expansion} from our general second-order theory of Eqs.~\eqref{eq:SIF_calculation_06} and \eqref{eq:SIF_calculation_07}. The second-order estimate $\KI^p$ writes as:
	\begin{equation}
		\label{eq:validation_general}
		\KI^p(\theta_1) = \KI^0(a_0,\theta_1) + \epsilon \KI^1(a_0, [\KI^0], [A_1]; \theta_1) + \epsilon^2 \left[\KI^1(a_0, [\KI^0], [A_2]; \theta_1) + \KI^2(a_0, [\KI^0], [A_1]; \theta_1) \right].
	\end{equation}
	Combining Eqs.~\eqref{eq:SIF_calculation_06} and \eqref{eq:validation_local_shapes}, one has
	\begin{equation}
		\label{eq:validation_cos_L}
		\begin{aligned}
			& \KI^1(a_0, [\KI^0], [A_1]; \theta_1) = \KI^0(a_0,\theta_1) \times \left[\dfrac{1}{4} - \dfrac{1}{2}\cos(2\theta_1)\right], \\
			\text{and } & \KI^1(a_0, [\KI^0], [A_2]; \theta_1) = \KI^0(a_0,\theta_1) \times \left[\dfrac{5}{32} - \dfrac{3}{16}\cos(2\theta_1) - \dfrac{1}{4}\cos(4\theta_1)\right].
		\end{aligned}	
	\end{equation}
	Similarly, the combination of Eqs.~\eqref{eq:SIF_calculation_07} and \eqref{eq:validation_local_shapes} yields:
	\begin{equation}
		\label{eq:validation_cos_H}
		\begin{aligned}
			\KI^2(a_0, [\KI^0], [A_1]; \theta_1) = \KI^0(a_0,\theta_1) \times \left[\dfrac{5}{32} - \dfrac{3}{16}\cos(2\theta_1) + \dfrac{1}{16}\cos(4\theta_1)\right].
		\end{aligned}	
	\end{equation}
	We observe that the two expansions coincide at $\theta = \theta_1$.
	
	\paragraph{Linear stress field dependent on $x$}~\\
	
	We consider next the case where the elliptical crack is loaded by a linear stress field dependent on $x$:
	\begin{equation}
		\label{eq:validation_sin_sigma}
		\sigma(z,x) = \sigma_0 \dfrac{x}{c} \Rightarrow \sigma(r,\theta) = \sigma_0 \dfrac{r}{c}\sin(\theta).
	\end{equation}
	As the procedure is similar to the previous case, we should be brief here. One has:
	\begin{equation}
		\label{eq:validation_sin_KI0}
		\KI^0(a,\theta) = \dfrac{4}{3\sqrt{\pi}}\dfrac{a^{3/2}}{c} \sigma_0 \sin(\theta).
	\end{equation}
	The exact SIF distribution along the elliptical crack front has been obtained by \citep{shah_stress_1971} and \citep{atroshchenko_stress_2009}. It reads:
	\begin{equation}
		\label{eq:validation_sin_KIe}
		\begin{aligned}
			&\KI^e(\varphi) = \sqrt{\pi c} \sigma_0\sin(\varphi) \dfrac{\kappa^2}{(1+\kappa^2)E(\kappa) - {\kappa'}^2K(\kappa)} \left((c/b)^4\cos^2(\varphi)+\sin^2(\varphi)\right)^{1/4} \\
			\Rightarrow\, & \KI^e(\theta) = \sqrt{\pi c} \sigma_0 \sin(\theta) \dfrac{\kappa^2}{(1+\kappa^2)E(\kappa) - {\kappa'}^2K(\kappa)} \dfrac{\left((c/b)^4\cos^2(\theta)+\sin^2(\theta)\right)^{1/4}}{\left((c/b)^2\cos^2(\theta)+\sin^2(\theta)\right)^{3/4}}. \\
		\end{aligned}	
	\end{equation}
	The expansion of $\KI^e/\KI^0$ at $\theta = \theta_1$ to second order in $\epsilon$ at $\epsilon=0$ and $\theta = \theta_1$ yields:
	\begin{equation}
		\label{eq:validation_sin_expansion}
		\begin{aligned}
			\KI^e(A_1)/\KI^0(a_0,\theta_1) & = \dfrac{3\pi}{4} \dfrac{\kappa^2}{(1+\kappa^2)E(\kappa) - {\kappa'}^2K(\kappa)} \left((c/b)^4\cos^2(\theta_1)+\sin^2(\theta_1)\right)^{1/4} \\
			& = 1 + \epsilon \left[-\dfrac{1}{4} - \dfrac{1}{2}\cos(2\theta_1)\right] + \epsilon^2 \left[\dfrac{1}{16} - \dfrac{1}{8}\cos(2\theta_1) - \dfrac{3}{16}\cos(4\theta_1)\right] + \mathcal{O}(\epsilon^3).
		\end{aligned}	
	\end{equation}
	
	Our goal is to retrieve again Eq.~\eqref{eq:validation_sin_expansion} from Eq.~\eqref{eq:validation_general}. One has:
	\begin{equation}
		\label{eq:validation_sin_L}
		\begin{aligned}
			& \KI^1(a_0, [\KI^0], [A_1]; \theta_1) = \KI^0(a_0,\theta_1) \times \left[-\dfrac{1}{4} - \dfrac{1}{2}\cos(2\theta_1)\right],\\
			\text{and } & \KI^1(a_0, [\KI^0], [A_2]; \theta_1) = \KI^0(a_0,\theta_1) \times \left[-\dfrac{3}{32} - \dfrac{5}{16}\cos(2\theta_1) - \dfrac{1}{4}\cos(4\theta_1)\right],
		\end{aligned}	
	\end{equation}
	and:
	\begin{equation}
		\label{eq:validation_sin_H}
		\begin{aligned}
			\KI^2(a_0, [\KI^0], [A_1]; \theta_1) = \KI^0(a_0,\theta_1) \times \left[\dfrac{5}{32} + \dfrac{3}{16}\cos(2\theta_1) + \dfrac{1}{16}\cos(4\theta_1)\right].
		\end{aligned}	
	\end{equation}
	The two expansions are found equal once again.
	
	\paragraph{Quadratic stress field odd in $z$ and $x$}~\\
	
	We conclude this Appendix with a last validation case, in which the stress field is quadratic and odd in both $z$ and $x$:
	\begin{equation}
		\label{eq:validation_sin2_sigma}
		\sigma(z,x) = \sigma_0 \dfrac{xy}{bc} \Rightarrow \sigma(r,\theta) = \sigma_0 \dfrac{r^2}{2bc}\sin(2\theta).
	\end{equation}
	For which one has:
	\begin{equation}
		\label{eq:validation_sin2_KI0}
		\KI^0(a,\theta) = \dfrac{8}{15\sqrt{\pi}}\dfrac{a^{5/2}}{bc} \sigma_0 \sin(2\theta).
	\end{equation}
	The exact SIF distribution along the elliptical crack front has been obtained by \citep{kassir_3D_1975} and \citep{atroshchenko_stress_2009}. It reads:
	\begin{equation}
		\label{eq:validation_sin2_KIe}
		\begin{aligned}
			&\KI^e(\varphi) = \sqrt{\pi c} \sigma_0\sin(2\varphi) \dfrac{\kappa^4}{2\left[2({\kappa'}^4-{\kappa'}^2+1)E(\kappa) - ({\kappa'}^4+{\kappa'}^2)K(\kappa)\right]} \left((c/b)^4\cos^2(\varphi)+\sin^2(\varphi)\right)^{1/4} \\
			\Rightarrow\, & \KI^e(\varphi) = \dfrac{\sqrt{\pi}c^{3/2}}{2b} \sigma_0 \sin(2\theta) \dfrac{\kappa^4}{2({\kappa'}^4-{\kappa'}^2+1)E(\kappa) - ({\kappa'}^4+{\kappa'}^2)K(\kappa)} \dfrac{\left((c/b)^4\cos^2(\theta)+\sin^2(\theta)\right)^{1/4}}{\left((c/b)^2\cos^2(\theta)+\sin^2(\theta)\right)^{5/4}}.
		\end{aligned}	
	\end{equation}
	The expansion of $\KI^e/\KI^0$ at $\theta = \theta_1$ to second order in $\epsilon$ at $\epsilon=0$ and $\theta = \theta_1$ then yields:
	\begin{equation}
		\label{eq:validation_sin2_expansion}
		\begin{aligned}
			\KI^e(A_1)/\KI^0(a_0,\theta_1) & = \dfrac{15\pi}{16} \dfrac{\kappa^4}{2({\kappa'}^4-{\kappa'}^2+1)E(\kappa) - ({\kappa'}^4+{\kappa'}^2)K(\kappa)} \left((c/b)^4\cos^2(\theta_1)+\sin^2(\theta_1)\right)^{1/4} \\
			& = 1 + \epsilon \left[- \dfrac{1}{2}\cos(2\theta_1)\right] + \epsilon^2 \left[\dfrac{3}{32} - \dfrac{1}{4}\cos(2\theta_1) - \dfrac{3}{16}\cos(4\theta_1)\right] + \mathcal{O}(\epsilon^3).
		\end{aligned}	
	\end{equation}
	We need now to compute the three different terms of our second-order equation~\eqref{eq:validation_general}. The terms arising from the linearization of $\KI$ yields:
	\begin{equation}
		\label{eq:validation_sin2_L}
		\begin{aligned}
			& \KI^1(a_0, [\KI^0], [A_1]; \theta_1) = \KI^0(a_0,\theta_1) \times \left[- \dfrac{1}{2}\cos(2\theta_1)\right],\\
			\text{and } & \KI^1(a_0, [\KI^0], [A_2]; \theta_1) = \KI^0(a_0,\theta_1) \times \left[-\dfrac{1}{16} - \dfrac{1}{4}\cos(2\theta_1) - \dfrac{1}{4}\cos(4\theta_1)\right],
		\end{aligned}	
	\end{equation}
	while the second-order contribution read:
	\begin{equation}
		\label{eq:validation_sin2_H1}
		\begin{aligned}
			\KI^2(a_0, [\KI^0], [A_1]; \theta_1) = \KI^0(a_0,\theta_1) \times \left[\dfrac{5}{32} + \dfrac{1}{16}\cos(4\theta_1)\right].
		\end{aligned}	
	\end{equation}
	The two expansions are found equal once more, although the shape of the stress field is more complex.
	
	\section{Ensemble averages of crack front propagating in weakly disordered materials}
	\label{apdx:statistics}
	
	This Appendix is dedicated to the statistical analysis of equation~\eqref{eq:disordered_normalized_toughness} defining the effective toughness of a random medium. In the following, we consider a statistical ensemble $\Omega$ containing all the possible realizations of the heterogeneous medium considered in Section~4. To each realization, we associate a specific real number $\omega$. We note $p~: \omega \mapsto p\left(\omega\right)$ the probability function associated to the ensemble $\Omega$. Thus, the probability that the variable $\omega'$ lies in some neighborhood of $\omega$ of measure $d\omega$ is $p(\omega)d\omega$. The mathematical expectation $E\left[u(x_1, x_2)\right]$ of any spatial observable $u: (x_1, x_2)\mapsto u(x_1, x_2)$ is defined as:
	\begin{equation}
		\average{u(x_1, x_2)} = \int_{\Omega} u(x_1, x_2 ; \omega) p(\omega) d\omega
	\end{equation}
	
	Using these notations, the ensemble averaging of Eq.~\eqref{eq:disordered_normalized_toughness} yields:
	\begin{equation}
		\label{eq:averages_KIc_eff}
		\average{\KIc^\mathrm{eff}/\KIc^0} = 1 + \bar{\sigma} L(0) \average{\hat{A}^1_0(a_0)} + \bar{\sigma}^2\left(L(0)\average{\hat{A}^2_0(a_0)} + H(0,0)\average{\hat{A}^1_0(a_0)^2} \right),
	\end{equation}
	so that its determination requires the prior calculation of $\average{\hat{A}^1_0(a_0)}$, $\average{\hat{A}^2_0(a_0)}$, and $\average{\hat{A}^1_0(a_0)^2}$. We start here with $\hat{A}^1_0(a_0)$, for which the calculation is the simplest and illustrates well the reasoning behind the ulterior ones. For an arbitrary mode $k$, $\average{\hat{A}^1_k(a_0)}$ reads:
	\begin{equation}
		\label{eq:averages_A1k}
		\begin{aligned}
			\average{\hat{A}^1_k(a_0)} & = \int_{\Omega} \hat{A}^1_k(a_0; \omega) p(\omega) d\omega = \int_{\Omega} \dfrac{\hat{f}_k(a_0; \omega)}{L(k)} p(\omega) d\omega = \int_{\Omega} \dfrac{1}{L(k)} \dfrac{1}{2\pi} \int_{0}^{2\pi} f(a_0, \theta; \omega) e^{-ik\theta} d\theta p(\omega) d\omega \\
			& = \dfrac{1}{2\pi L(k)} \int_{0}^{2\pi} \underset{=0}{\average{f(a_0, \theta; \omega)}} e^{-ik\theta} d\theta = 0,
		\end{aligned}	
	\end{equation}
	where we used the expression of $\hat{A}^1_k(a_0)$ given in Eq.~\eqref{eq:disordered_A1_fourier}. Even if only the value of $\average{\hat{A}^1_0(a_0)}$ to estimate the average value of the effective toughness, our calculations show that:
	\begin{equation}
		\label{eq:averages_A1}
		\average{A_1(a_0, \theta)} = \average{\sum_k \hat{A}^1_k(a_0) e^{ik\theta}} = \sum_k \average{\hat{A}^1_k(a_0)} e^{ik\theta} = 0,
	\end{equation}
	meaning that the first-order contribution to the crack front deformations are on average zero.\\
	
	We proceed similarly for $A_2$, and start by averaging Eq.~\eqref{eq:disordered_A2_fourier}:
	\begin{equation}
		\label{eq:averages_A2k_01}
		\average{\hat{A}^2_k(a_0)} = \dfrac{1}{L(k)} \sum_{k'} \left(\dfrac{a_0}{L(k-k')} \average{\widehat{\left(\dfrac{\partial f}{\partial r}\right)}_{k'}\,(a_0) \hat{f}_{k-k'}(a_0)} - \dfrac{H(k', k-k')}{L(k')L(k-k')}\average{\hat{f}_{k'}(a_0)\hat{f}_{k-k'}(a_0)}\right)
	\end{equation}
	Let us first concentrate on:
	\begin{equation}
		\label{eq:averages_fkfk_01}
		\average{\hat{f}_{k}(a_0)\hat{f}_{k'}(a_0)} = \dfrac{1}{4\pi^2} \int_{0}^{2\pi}\int_{0}^{2\pi} \average{f(a_0, \theta)f(a_0, \theta')} e^{-i(k\theta+k'\theta')} d\theta d\theta'
	\end{equation}
	To calculate Eq.~\eqref{eq:averages_fkfk_01}, it is convenient to switch to Cartesian coordinates. The pairs $(r,\theta)$ and $(r',\theta')$ are then equivalent to$(z=a_0\cos\theta, x=a_0\sin\theta)$ and $(z'=a_0\cos\theta', x'=a_0\sin\theta')$. With these notations, one has:
	\begin{equation}
		\label{eq:averages_ff}
		\begin{aligned}
			\average{f(a_0, \theta)f(a_0, \theta')} & = \average{f^*(a_0\cos\theta, a_0\sin\theta)f^*(a_0\cos\theta', a_0\sin\theta'} e^{-i(k\theta+k'\theta')} \\ 
			&= \mathcal{F}_z\left(z-z'\right) \mathcal{F}_x\left(x-x'\right)\\
			& = e^{-4(a_0/d)^2\sin^2[(\theta-\theta')/2]}.
		\end{aligned}
	\end{equation}
	where we used the expressions of $\mathcal{F}_z$ and $\mathcal{F}_x$ given in Eq.~\eqref{eq:disordered_correlations_shape}, as well as the following identities: 
	\begin{equation*}
		\begin{cases}
			z-z' = -2a_0\sin[(\theta-\theta')/2]\sin[(\theta+\theta')/2]\\
			x-x' = 2a_0\sin[(\theta-\theta')/2]\cos[(\theta+\theta')/2].
		\end{cases}
	\end{equation*}
	Injecting Eq.~\eqref{eq:averages_ff} into \eqref{eq:averages_fkfk_01}, one finds:
	\begin{equation}
		\label{eq:averages_fkfk_02}
		\begin{aligned}
			\average{\hat{f}_{k}(a_0)\hat{f}_{k'}(a_0)} & = \dfrac{1}{4\pi^2} \int_{0}^{2\pi}\int_{0}^{2\pi} e^{-4(a_0/d)^2\sin^2[(\theta-\theta')/2]} e^{-ik(\theta-\theta')} e^{-i(k+k')\theta'} d\theta d\theta'
			& = \Delta_{-kk'} \hat{F}_k(a_0/d),
		\end{aligned}
	\end{equation}
	where $ \hat{F}_k$ is the $k$-th Fourier coefficient of the $2\pi$-periodic function:
	\begin{equation}
		\label{eq:averages_function_F}
		F(a_0/d, \theta) = e^{-4(a_0/d)^2\sin^2(\theta/2)},
	\end{equation}
	which relates to the correlation shape $\mathcal{F}$ of Eq.~\eqref{eq:disordered_correlations_shape}. We must now calculate the second average term in the sum of Eq.\eqref{eq:averages_A2k_01}. It reads:
	\begin{equation}
		\label{eq:averages_dfkfk_01}
		\average{\widehat{\left(\dfrac{\partial f}{\partial r}\right)}_{k}\hat{f}_{k'}(a_0)} = \dfrac{1}{4\pi^2} \int_{0}^{2\pi}\int_{0}^{2\pi} \average{\dfrac{\partial f}{\partial r}(a_0, \theta)f(a_0, \theta')} e^{-i(k\theta+k'\theta')} d\theta d\theta'
	\end{equation}
	The calculation of equation~\eqref{eq:averages_dfkfk_01} proves more difficult than that of Eq.~\eqref{eq:averages_ff}. In the following, use will be made of Fourier transforms. The definition adopted here for the single Fourier transform $\tilde{g}(p)$ of an arbitrary function $g(x)$ is:
	\begin{equation}
		\label{eq:single_fourier_transform}
		\tilde{g}(p) = \int_{-\infty}^{+\infty} g(x) e^{-ipx} dx \Longleftrightarrow g(x) = \dfrac{1}{2\pi} \int_{-\infty}^{+\infty} \tilde{g}(p) e^{ipx} dp.
	\end{equation}
	and the double $(z,x)$ Fourier transform $\dbtilde{h}(m,n)$ of an arbitrary function $h(z,x)$ reads:
	\begin{equation}
		\label{eq:double_fourier_transform}
		\dbtilde{h}(m,n) = \int_{-\infty}^{+\infty}\int_{-\infty}^{+\infty} h(z,x) e^{-i(mz+nx)} dz dx \Longleftrightarrow h(z,x) = \frac{1}{4\pi^2}\int_{-\infty}^{+\infty}\int_{-\infty}^{+\infty} \dbtilde{h}(m,n) e^{i(mz+nx)} dm dn
	\end{equation}
	One then has:
	\begin{equation}
		\label{eq:averages_dff_01}
		\begin{aligned}
			\average{\dfrac{\partial f}{\partial r}(a_0, \theta)f(a_0, \theta')} & = \int_{\Omega} \dfrac{\partial f}{\partial r}(a_0, \theta; \omega)f(a_0, \theta'; \omega) p(\omega) d\omega \\
			& = \int_{\Omega} \left[\dfrac{\partial f^*}{\partial z}(z,x; \omega) \cos\theta + \dfrac{\partial f^*}{\partial x}(z,x; \omega) \sin\theta\right] f^*(z', x'; \omega) p(\omega) d\omega \\
			& = \dfrac{1}{(2\pi)^4} \int_{m, n, m', n'} i(m\cos\theta + n\sin\theta) \average{\tilde{f^*}(m,n) \tilde{f^*}(m',n')} e^{i(mz+nx+m'z'+n'x')} dm dn dm'dn'.
		\end{aligned}
	\end{equation}
	As shown by \cite{favier_statistics_2006} in their Appendix B, one has:
	\begin{equation}
		\label{eq:averages_TfTf}
		\begin{aligned}
			\average{\tilde{f^*}(m,n) \tilde{f^*}(m',n')} & = (2\pi)^2 \delta(m+m') \delta(n+n') \widetilde{\mathcal{F}}_z(m) \widetilde{\mathcal{F}}_x(n),
		\end{aligned}
	\end{equation}
	where $\delta$ is the Dirac function, and $\widetilde{\mathcal{F}}_z$ and $\widetilde{\mathcal{F}}_x$ are the Fourier transforms of the correlation functions $\mathcal{F}_z$ and $\mathcal{F}_x$. Injecting Eq.~\eqref{eq:averages_TfTf} into \eqref{eq:averages_dff_01}, one finds:
	\begin{equation}
		\label{eq:averages_dff_02}
		\begin{aligned}
			\average{\dfrac{\partial f}{\partial r}(a_0, \theta)f(a_0, \theta')} & = \dfrac{1}{(2\pi)^2} \int_{m, n} i(m\cos\theta + n\sin\theta) \widetilde{\mathcal{F}}_z(m) \widetilde{\mathcal{F}}_x(n) e^{i(m(z-z')+n(x-x'))} dm dn dm'dn'\\
			& = \mathcal{F}'_z(z-z')\mathcal{F}_x(x-x')\cos\theta + \mathcal{F}_z(z-z')\mathcal{F}'_x(x-x') \sin\theta\\
			& = - 4a_0/d^2 \sin^2[(\theta-\theta')/2] e^{-4(a_0/d)^2\sin^2[(\theta-\theta')/2]}
		\end{aligned}
	\end{equation}
	where we used the following identities:
	\begin{equation*}
		\begin{cases}
			\mathcal{F}'_z(z-z')\mathcal{F}_x(x-x') = 4a_0/d^2 \sin[(\theta-\theta')/2]\sin[(\theta+\theta')/2] e^{-4(a_0/d)^2\sin^2[(\theta-\theta')/2]} \\
			\mathcal{F}_z(z-z')\mathcal{F}'_x(x-x') = - 4a_0/d^2 \sin[(\theta-\theta')/2]\sin[(\theta+\theta')/2] e^{-4(a_0/d)^2\sin^2[(\theta-\theta')/2]},
		\end{cases}
	\end{equation*}
	and:
	\begin{equation*}
		\begin{cases}
			\cos\theta = \cos[(\theta-\theta')/2]\cos[(\theta+\theta')/2] - \sin[(\theta-\theta')/2]\sin[(\theta+\theta')/2] \\
			\sin\theta = \sin[(\theta-\theta')/2]\cos[(\theta+\theta')/2] + \cos[(\theta-\theta')/2]\sin[(\theta+\theta')/2].
		\end{cases}
	\end{equation*}
	The combination of Eqs.~\eqref{eq:averages_dfkfk_01} and \eqref{eq:averages_dff_02} finally yields:
	\begin{equation}
		\label{eq:averages_dfkfk_02}
		\begin{aligned}
			\average{\widehat{\left(\dfrac{\partial f}{\partial r}\right)}_{k}\hat{f}_{k'}(a_0)} & = \dfrac{1}{4\pi^2} \int_{0}^{2\pi}\int_{0}^{2\pi} - 4(a_0/d)^2 \sin^2[(\theta-\theta')/2] e^{-4(a_0/d)^2\sin^2[(\theta-\theta')/2]} e^{-ik(\theta-\theta')} e^{-i(k+k')\theta'} d\theta d\theta' \\
			& = \Delta_{-kk'} \hat{G}_k(a_0/d)/a_0,
		\end{aligned}
	\end{equation}
	where $ \hat{G}_k$ is the $k$-th Fourier coefficient of the $2\pi$-periodic function:
	\begin{equation}
		\label{eq:averages_function_G}
		G(a_0/d, \theta) = -4(a_0/d)^2\sin^2(\theta/2) e^{-4(a_0/d)^2\sin^2(\theta/2)}
	\end{equation}
	
	As a result, all $\average{\hat{A}^2_k(a_0)}$ are zero, except for $k=0$, for which one gets:
	\begin{equation}
		\label{eq:averages_A2k}
		\average{\hat{A}^2_0(a_0)} = \dfrac{1}{L(0)} \sum_{k'} \left(\dfrac{1}{L(k')} \hat{G}_{k'}(a_0/d)- \dfrac{H(k',-k')}{L(k')^2} \hat{F}_{k'}(a_0/d) \right)
	\end{equation}
	This means that the average second-order contribution to the front deformations is a simple dilation of the reference circular front of radius $a_0$ by a quantity:
	\begin{equation}
		\label{eq:averages_A2}
		\average{A_2(a_0, \theta)} = \dfrac{1}{L(0)} \sum_{k'} \left(\dfrac{1}{L(k')} \hat{G}_{k'}(a_0/d) - \dfrac{H(k',-k')}{L(k')^2} \hat{F}_{k'}(a_0/d) \right)
	\end{equation}
	We only need to calculate the third and last missing term of Eq.~\eqref{eq:averages_KIc_eff}. Using Eq.~\eqref{eq:averages_fkfk_02}, one has:
	\begin{equation}
		\label{eq:averages_A10_square}
		\average{\hat{A}^1_0(a_0)^2} = \dfrac{\average{\hat{f}_0(a_0)^2}}{L(0)^2} = \dfrac{\hat{F}_0(a_0/d)}{L(0)^2},
	\end{equation}
	so that the average of the effective toughness finally reads:
	\begin{equation}
		\label{eq:averages_effective_toughness}
		\average{\KIc^\mathrm{eff}/\KIc^0} = 1 + \bar{\sigma}^2 \left[\sum_{k'} \left(\dfrac{1}{L(k')} \hat{G}_{k'}(a_0/d) - \dfrac{H(k',-k')}{L(k')^2} \hat{F}_{k'}(a_0/d)\right) + \dfrac{H(0,0)}{L(0)^2}\hat{F}_0(a_0/d) \right] + \mathcal{O}(\bar{\sigma}^3).
	\end{equation}
	One can additionally show that: 
	\begin{equation}
		\label{eq:variance_effective_toughness}
		\begin{aligned}
			\Var{\KIc^\mathrm{eff}/\KIc^0} & = \average{(\KIc^\mathrm{eff}/\KIc^0)^2} - \average{\KIc^\mathrm{eff}/\KIc^0}^2 = \bar{\sigma}^2 L(0)^2 \average{\hat{A}^1_0(a_0)^2} \\
			& = \bar{\sigma}^2 \hat{F}_0(a_0/d) + \mathcal{O}(\bar{\sigma}^3).
		\end{aligned}
	\end{equation}
	
	\bibliographystyle{elsarticle-harv}

\end{document}